\documentclass[12pt]{article}
\usepackage{graphicx}

\def\Re{{\cal R \mskip-4mu \lower.1ex \hbox{\it e}\,}}
\def\Im{{\cal I \mskip-5mu \lower.1ex \hbox{\it m}\,}}
\def\ie{{\it i.e.}}
\def\eg{{\it e.g.}}

\def\etal{{\it et al.}}

\def\sub#1{_{\lower.25ex\hbox{$\scriptstyle#1$}}}
\def\tev{\,{\ifmmode\mathrm {TeV}\else TeV\fi}}
\def\gev{\,{\ifmmode\mathrm {GeV}\else GeV\fi}}
\def\mev{\,{\ifmmode\mathrm {MeV}\else MeV\fi}}
\def\mpl{\ifmmode M_{pl}\else $M_{pl}$\fi}
\def\mpl{\ifmmode \overline M_{Pl}\else $\bar M_{Pl}$\fi}
\def\to{\rightarrow}

\def\subw{_{\rm w}}
\def\mh{\ifmmode m\sbl H \else $m\sbl H$\fi}
\def\mch{\ifmmode m_{H^\pm} \else $m_{H^\pm}$\fi}
\def\mt{\ifmmode m_t\else $m_t$\fi}
\def\mc{\ifmmode m_c\else $m_c$\fi}
\def\mz{\ifmmode M_Z\else $M_Z$\fi}
\def\mw{\ifmmode M_W\else $M_W$\fi}
\def\mws{\ifmmode M_W^2 \else $M_W^2$\fi}
\def\mhs{\ifmmode m_H^2 \else $m_H^2$\fi}   
\def\mzs{\ifmmode M_Z^2 \else $M_Z^2$\fi}
\def\mts{\ifmmode m_t^2 \else $m_t^2$\fi}
\def\mcs{\ifmmode m_c^2 \else $m_c^2$\fi}
\def\mchs{\ifmmode m_{H^\pm}^2 \else $m_{H^\pm}^2$\fi}
\def\ztwo{\ifmmode Z_2\else $Z_2$\fi}
\def\zone{\ifmmode Z_1\else $Z_1$\fi}
\def\mtwo{\ifmmode M_2\else $M_2$\fi}
\def\mone{\ifmmode M_1\else $M_1$\fi}
\def\tb{\ifmmode \tan\beta \else $\tan\beta$\fi}
\def\xw{\ifmmode x\subw\else $x\subw$\fi}
\def\ch{\ifmmode H^\pm \else $H^\pm$\fi}
\def\lum{\ifmmode {\cal L}\else ${\cal L}$\fi}
\def\inpb{\,{\ifmmode {\mathrm {pb}}^{-1}\else ${\mathrm {pb}}^{-1}$\fi}}
\def\infb{\,{\ifmmode {\mathrm {fb}}^{-1}\else ${\mathrm {fb}}^{-1}$\fi}}
\def\epem{\ifmmode e^+e^-\else $e^+e^-$\fi}
\def\ppb{\ifmmode \bar pp\else $\bar pp$\fi}
\def\bsg{\ifmmode B\to X_s\gamma\else $B\to X_s\gamma$\fi}
\def\bsll{\ifmmode B\to X_s\ell^+\ell^-\else $B\to X_s\ell^+\ell^-$\fi}
\def\bstt{\ifmmode B\to X_s\tau^+\tau^-\else $B\to X_s\tau^+\tau^-$\fi}
\def\lamt{\ifmmode \tilde\lambda\else $\tilde\lambda$\fi}
\def\shat{\ifmmode \hat s\else $\hat s$\fi}
\def\that{\ifmmode \hat t\else $\hat t$\fi}
\def\uhat{\ifmmode \hat u\else $\hat u$\fi}

\newskip\zatskip \zatskip=0pt plus0pt minus0pt
\def\matth{\mathsurround=0pt}
\def\lsim{\mathrel{\mathpalette\atversim<}}

\def\atversim#1#2{\lower0.7ex\vbox{\baselineskip\zatskip\lineskip\zatskip
  \lineskiplimit 0pt\ialign{$\matth#1\hfil##\hfil$\crcr#2\crcr\sim\crcr}}}

\def\grtsim{\,\,\rlap{\raise 3pt\hbox{$>$}}{\lower 3pt\hbox{$\sim$}}\,\,}
\def\lsim{\,\,\rlap{\raise 3pt\hbox{$<$}}{\lower 3pt\hbox{$\sim$}}\,\,}

\renewcommand{\thefootnote}{\fnsymbol{footnote}}

\begin{document} \begin{titlepage}
\rightline{\vbox{\halign{&#\hfil\cr
&SLAC-PUB-11036\cr
&March 2005\cr}}}
\begin{center}
\thispagestyle{empty} \flushbottom { {\Large\bf Collider 
Production of TeV Scale Black Holes and Higher-Curvature Gravity 
\footnote{Work supported in part
by the Department of Energy, Contract DE-AC02-76SF00515}
\footnote{e-mail:
$^a$rizzo@slac.stanford.edu}}}
\medskip
\end{center}

\centerline{Thomas G. Rizzo$^{a}$}
\vspace{8pt} 
\centerline{\it Stanford Linear
Accelerator Center, 2575 Sand Hill Rd., Menlo Park, CA, 94025}

\vspace*{0.3cm}

\begin{abstract}
We examine how the production of TeV scale black holes at colliders is 
influenced by the presence of Lovelock higher-curvature terms in the  
action of models with large extra dimensions. Such terms are expected to 
arise on rather general grounds, \eg, from string theory and are often used 
in the literature to model modifications to the Einstein-Hilbert action 
arising from quantum and/or stringy corrections. While adding the invariant  
which is quadratic in the curvature leads to quantitative modifications in 
black hole properties, cubic and higher invariants are found to produce 
significant qualitative changes, \eg, classically stable black holes. We use 
these higher-order curvature terms to construct a toy model of the black hole 
production cross section threshold. For reasonable parameter values we 
demonstrate that detailed measurements of the properties of black holes at 
future colliders will be highly sensitive to the presence of the Lovelock 
higher-order curvature terms. 
\end{abstract}



\renewcommand{\thefootnote}{\arabic{footnote}} \end{titlepage} 

%
%
%
%

\section{Introduction}

The introduction of large extra dimensions by Arkani-Hamed, Dimopoulos and 
Dvali{\cite {ADD}} offers us the possibility that the true fundamental
scale of gravity, $M_*$, may not be too far above the weak scale, $\sim$ TeV. 
In this scenario, gravity is allowed to propagate in all $D=4+n$ dimensions 
while the Standard Model(SM) fields are confined to live on a 
three-dimensional 'brane' which we assume to be flat. One then 
finds $M_*$ is related to the usual 
4-d (reduced) Planck scale, $\mpl$, via the now famous expression 
\begin{equation}
\mpl^2=V_nM_*^{n+2}\,,
\end{equation}
where $V_n$ is the volume of the compactified extra dimensions. These 
compactified dimensions are usually assumed to be flat, \ie, they form an 
$n$-dimensional torus so that, if all compactification radii, $R_c$, were the 
same, $V_n=(2\pi R_c)^n$. This simple 
ADD picture has three basic predictions that have gotten significant 
attention in the literature{\cite {JM}} over the past few years: ($i$) the 
emission of 
graviton Kaluza-Klein states during the collision of SM particles leading to 
final states with apparent missing energy{\cite {GRW,HLZ,PP}}; ($ii$) the 
exchange of graviton Kaluza-Klein excitations between SM fields 
leading to new dimension-8 
contact interaction-like  operators with distinctive spin-2 
properties{\cite {GRW,HLZ,JLH}; ($iii$) the 
production of black holes(BH) at colliders and in cosmic rays 
with geometric subprocess cross sections once 
energies greater than $\sim M_*$ are exceeded{\cite {Fisch,DL,GT,Kanti}}. 
While ($i$) and ($ii$) result 
from an expansion of the gravitational action, \ie, the $D-$dimensional 
Einstein-Hilbert(EH) action, to leading order in the gravitational field and 
are in some sense perturbative, ($iii$) on the other-hand relies upon the full 
non-perturbative content of the EH action. Here one is really testing 
$D-$dimensional General Relativity and not {\it just} the ADD picture. 
Within the ADD scenario, collider measurements of ($i$) and ($ii$) type 
processes will tell us the values of $n$ and $M_*${\cite {JM}}. 

If anything like the ADD scenario is realized in Nature it will have to 
be part of a much larger framework, \ie, ADD is at best an effective theory 
that operates at energies below the scale $M_*$. It is reasonable to expect 
that at least some aspects of this more complete theory may leak down into 
the collider tests listed above and may lead to significant quantitative 
and/or qualitative 
modifications that can be probed experimentally. Here we examine one such 
extension of the basic model:  the existence of higher curvature terms 
which augment the EH action. Such terms are expected to be present at some 
level on rather general grounds from string 
theory{\cite {Zwiebach,Mavromatos}} or other possible 
high-scale completions of General Relativity. These terms can also be 
thought of as being generated by quantum corrections to the ordinary EH 
action. Given arbitrary powers and 
derivatives of the curvature tensor there are a huge number of possible 
higher-order invariants from which to choose. In order to decide what 
to add to the EH action we need some form of guidance in making possible 
selections.
 
Fortunately, a certain special class of such invariants with very 
interesting properties was first generally described long ago  
by Lovelock{\cite {Lovelock}} and, hence, are termed Lovelock invariants.
These are constructed out of powers of the curvature tensor with no additional 
derivatives. They are also sometimes referred to in the literature as 
generalized Euler densities since their volume integrals are related to the 
Euler characteristics.  These Lovelock invariants themselves come in fixed 
order, $m$, which we denote here as ${\cal L}_m$,  
that describes the number of powers of the curvature tensor, contracted in 
various ways, out of which they are constructed. Apart from normalization 
factors we can express the ${\cal L}_m$ as 
\begin{equation}
{\cal L}_m \sim \delta^{A_1B_1...A_mB_m}_{C_1D_1...C_mD_m}~R_{A_1B_1}
~^{C_1D_1}.....R_{A_mB_m}~^{C_mD_m}\,,
\end{equation}
where $\delta^{A_1B_1...A_mB_m}_{C_1D_1...C_mD_m}$ is the totally 
antisymmetric product of Kronecker deltas and $R_{AB}~^{CD}$ is the 
$D$-dimensional curvature tensor. Fortunately, as can be seen 
by this definition, the 
number of such invariants that can exist in any given dimension is highly 
constrained. For a space with an even number of dimensions, $D=2m$, the 
Lovelock invariant is a topological one and leads to a total derivative, \ie, 
a surface term, in the action.  All of the higher order invariants, 
$D\leq 2m-1$, can then be shown to vanish identically by various curvature 
tensor index symmetry properties. On the otherhand, for the cases with 
$D\geq 2m+1$, the ${\cal L}_m$ are true dynamical objects that once 
added the action can significantly alter the field equations normally 
associated with the EH term. However it can be shown that the addition of 
any or all of the ${\cal L}_m$ to the EH action still results in a theory 
with only second order equations of motion as is the case 
for ordinary Einstein gravity. Furthermore, variation of the new action 
leads to modifications of Einstein's equations by the addition of 
new terms which are second-rank symmetric tensors with vanishing covariant 
derivatives and which depend only on the metric and its first and second 
derivatives, \ie, they have the same general properties as the Einstein 
tensor itself but are higher order in the curvature. 
These properties are quite special. Generally, the addition to the action 
of arbitrary 
invariants formed from ever higher powers of the curvature tensor will lead to 
equations of motion of ever higher order, \ie, ever more co-ordinate 
derivatives of the metric tensor and graviton field, \eg, terms with quartic 
derivatives. Such theories will have very serious problems with both the  
presence of ghosts as well as with unitarity{\cite {Zwiebach}}. The 
Lovelock invariants are constructed in such a way as to be free of these 
problems making them quite exceptional. Interestingly the Lovelock invariants 
are found to be just the forms taken by the higher order curvature terms 
generated in perturbative critical 
string theory{\cite {Zwiebach,Mavromatos}}. This is 
perhaps what we might have expected if string theories are to avoid these 
troublesome ghost and unitarity issues. 

In the case of the ADD model, we are generally considering the possibility 
that the number of extra dimensions lies in the range 
$2\leq n\leq 6$.{\footnote {Note that the case $n=1$ is excluded by 
Solar System measurements and the value $n=2$ with a low $M_*$ 
is somewhat disfavored by 
astrophysical constraints{\cite {JM}}.}} We now imagine extending the 
$D$-dimensional EH action of ADD to include all of the allowable Lovelock 
invariants for a given value of $n$ 
with arbitrary coefficients. What do these additional terms do? 
Firstly, as we will see below based on simple dimensional analysis, 
these new invariants are suppressed 
in comparison to the usual EH action by powers of $M_*$ and so are only 
important as TeV-scale energies are approached. Secondly, a short amount 
of algebra demonstrates that their impact on the ADD model itself 
is rather subtle since the compactified space is assumed to be flat and 
the SM fields are restricted to a 3-brane. In particular the leading-order 
expression for the interaction of the graviton Kaluza-Klein modes, 
$h_{\mu\nu}^{(n)}$ with the localized SM fields remains as in the EH case:
\begin{equation}
S_{int}=\int d^4x~d^ny~ {-1\over {M_*^{1+n/2}}}\sum_n h_{\mu\nu}^{(n)}
T^{\mu\nu}\delta(y)\,,
\end{equation}
where $T^{\mu\nu}$ is the stress-energy tensor for the SM fields and with 
$y$ labeling the coordinates of the compactified torus. (Graviton 
self-interactions, not generally probed in the ADD scenario {\it are} 
modified by the Lovelock terms.) The mass 
spectrum of the graviton KK states is also unchanged. This means that 
the signals ($i$) and ($ii$) remain unaltered by these modifications to the 
ADD action. Note that if the geometry of space was {\it curved}, as in the 
Randall-Sundrum model{\cite {RS}}, the couplings of graviton Kaluza-Klein 
excitations to SM matter would be altered{\cite {TGR}} by the presence of the  
${\cal L}_m$ as would the KK spectrum. 
It is interesting to note that in the RS case we  
observe that as the curvature parameter $k/\mpl\to 0$ we find that the 
matter-graviton couplings return to those of usual RS scenario even though 
${\cal L}_2$ is present with a non-zero coefficient. 
What about ($iii$), the production of BH at colliders? This is the subject 
of the current paper as presented in our discussion below. 

The outline of the paper is as follows: In Section 2 we will present the 
basic background and formalism associated with the Lovelock invariants and 
show how they would alter EH expectations for the Schwarzschild radius, 
temperature, entropy, free energy and heat capacity/specific heat of BH. 
In Section 3 we 
will make a numerical analysis of the Lovelock induced modifications for TeV 
scale BH and determine how their anticipated properties at 
colliders would be affected. In particular we will show 
that while the presence of ${\cal L}_2$ can quantitatively alter the details 
of the properties of BH, the presence of ${\cal L}_{3,4}$ terms may lead to 
significant qualitative modifications in their properties, \eg, the existence 
of classically stable BH with minimum allowed masses and corresponding mass 
thresholds. In Section 4, based on the analysis in the previous sections,  
we present a toy model for the threshold behavior of the BH production 
cross section at colliders in odd 
dimensions. Section 5 contains a discussion and our conclusions.

\section{Formalism}

Before beginning the general discussion, 
the first question one might ask 
is `Just how many Lovelock invariants are there for our cases of 
interest?' For 
simplicity, let us begin with 4-d; apart from 
numerical factors, only ${\cal L}_0$, which we can think of as the 
cosmological constant, and ${\cal L}_1=R$, the usual Ricci scalar of EH, can 
be present and dynamical. The invariant of the next order, 
${\cal L}_2$, can be identified with the Gauss-Bonnet(G-B)  
invariant, $R^2-4R_{\mu\nu}R^{\mu\nu}+
R_{\mu\nu\alpha\beta}R^{\mu\nu\alpha\beta}$, which is a topological term 
as well as a total derivative and can be written {\it in 4-d} as 
$\epsilon^{\mu\nu\rho\sigma}\epsilon_{\alpha\beta\gamma\delta}
R_{\mu\nu}~^{\alpha\beta}R_{\rho\sigma}~^{\gamma\delta}$. All 
higher order invariants can be shown to vanish. 
Thus the most general Lovelock theory in 
4-d is just EH plus a cosmological constant which is just ordinary General 
Relativity. Now in 5-d, all of the 
${\cal L}_{m \geq 3}$ still vanish as in 4-d but the G-B invariant 
is no longer a total derivative and its presence will modify the results 
obtained from Einstein gravity. The generalization is now quite clear: 
for $D=5,6$ only ${\cal L}_{0-2}$ can be present. For $D=7,8$ only 
${\cal L}_{0-3}$ can be present while for $D=9,10$ only ${\cal L}_{0-4}$. 
Since ADD assumes that the compactified space is flat the coefficient of 
${\cal L}_{0}$ is taken to be zero and, to reproduce conventional results,  
${\cal L}_{1}$ is normalized so that it can be 
identified with the usual EH term. 
Thus there are at most three new pieces to add to the EH action and so  
the general form for the extended ADD model we consider is given by 
\begin{equation}
S=\int d^{4+n}x ~\sqrt {-g}~{M_*^{n+2}\over {2}} \Bigg[R+
{\alpha\over {M_*^2}}{\cal L}_2+{\beta \over {M_*^4}}{\cal L}_3+
{\gamma \over {M_*^6}}{\cal L}_4\Bigg]\,,
\end{equation}
where $\alpha$, $\beta$ and $\gamma$ are dimensionless coefficients which we 
take to be positive in our discussion below. (The size of these coefficients 
will be discussed later.) It is important to note that the fundamental 
mass parameter, $M_*$ appearing in the above action is the {\it same} as the 
one appearing in the ADD $M_*-\mpl$ relationship Eq.(1) and in the coupling of 
the SM brane fields to the graviton excitations Eq.(2). 

Note that $M_*$ can be related to several other mass parameters used in the 
literature. The Planck scale employed by 
Dimopoulos and Landsberg{\cite {DL}} is 
given by $M_{DL}=(8\pi)^{1/(n+2)}M_*$ while that of Giddings and 
Thomas{\cite {GT}} is found to be $M_{GT}=[2(2\pi)^n]^{1/(n+2)}M_*$;  
moreover, Giudice \etal {\cite {GRW}} employ a different scale 
$M_D=(2\pi)^{n/(n+2)}M_*$. Further note that $M_*$ is thus correspondingly 
{\it smaller} than all of these other 
parameters with consequently far weaker bounds{\cite {JM}}. For example, if 
$n=2(6)$ and $M_D=1.5$ TeV then $M_*=0.60(0.38)$ TeV; collider bounds on 
$M_*$ are thus well below 1 TeV at present. The advantage 
of employing $M_*$ is 
that it a common parameter occurring 
not only in the original action but in the 
resulting SM matter couplings to gravity 
as well as in the important ADD relationship Eq.(1). 

As discussed above, in $D=4+n$ dimensions 
${\cal L}_2$ is given by the generalized Gauss-Bonnet form: 
\begin{equation}
{\cal L}_2=R^2-4R_{AB}R^{AB}+R_{ABCD}R^{ABCD}\,,
\end{equation}
while for ${\cal L}_3$, first given explicitly by 
M\"uller-Hoisson{\cite {MH}, one has
\begin{eqnarray}
{\cal L}_3&=&R^3-12RR^{AB}R_{AB}+2R^{ABCD}R_{CDEF}R^{EF}~_{AB}+8R^{AB}~_{CE}
R^{CD}~_{BF}R^{EF}~_{AD}\nonumber \\
&+&24R^{ABCD}R_{CDBE}R^E~_A+3RR^{ABCD}R_{CDAB}+24R^{ACBD}R_{BA}R_{DC}
\nonumber \\
&+&16 R^{AB}R_{BC}R^C~_A\,.
\end{eqnarray}
${\cal L}_4$, which has 25 terms, was first given explicitly by 
Briggs{\cite {Briggs}} and will not be reproduced here but the general 
qualitative pattern is now obvious. As discussed above the Lovelock invariants 
lead to an augmented set of Einstein's equations of the form 
\begin{equation}
R_{AB}-{1\over {2}}g_{AB}+{{2\alpha}\over {M_*^2}}L^{(2)}_{AB}+....
={1\over {M_*^{n+2}}}
T_{AB}\,,
\end{equation}
with, \eg, $L^{(2)}_{AB}$  given by 
\begin{equation}
L^{(2)}_{AB} = RR_{AB}-R_{ACGH}R^{GHC}~_B-2R_{AGBH}R^{HG}-2R_{AC}R^C~_B+
{1\over {4}}g_{AB}{\cal L}_2\,.
\end{equation}
As mentioned above the set of tensors $L^{(m)}_{AB}$ are symmetric and have 
vanishing covariant derivatives.

The possibility that TeV scale black holes(BH) may be a copious signal for 
extra dimensions at future colliders has been discussed by a number of 
authors{\cite {DL,GT}}; the corresponding production by cosmic rays has been 
considered as well, \eg, in Ref.{\cite {Anchor}}. 
A first step in any consideration of this process is to ask just how large 
the cross section can be. 
A leading approximation for the subprocess cross-section 
for the production of a {\it Schwarzschild} BH of mass $M_{BH}$ is 
simply given by the geometric BH size{\cite {GR,Rychkov,Kanti}} once we reach  
energies in excess of $M_*$, \ie, 
\begin{equation}
\hat \sigma \simeq \pi R^2 ~\theta(\sqrt s-M_*^2)\,,
\end{equation}
with $s=M_{BH}^2$ and where from this point forward $R$ is the 
$(4+n)$-dimensional Schwarzschild radius corresponding to 
the mass $M_{BH}$. Notice that the rather unphysical threshold behavior 
of this cross section is just a simple step function. While an understanding 
of this kinematic threshold regime 
is likely beyond any semi-classical approach we can be sure that the 
true threshold will not be so trivial. It is now believed that this 
cross section 
expression correctly describes the BH production process for $M_{BH}>>M_*$ 
up to an overall factor of order unity{\cite {GR}} within $D$-dimensional 
General Relativity based on the EH action. In fact, using this expression the 
production of BH at the LHC has been studied in detail, including some 
detector 
effects, in Refs.{\cite {Harris,Tanaka}}. These authors have shown how it 
is possible to extract information about the values of $M_{BH}$, $n$, and 
$M_*$ from LHC data, once BH production is observed, owing to the very large 
statistics that is expected. 

The usual description of the BH 
production process assumes that the initial horizon forms around a BH state 
which is in general 
both charged and rapidly rotating due to the initial large impact 
parameter involved in the collision and the nature of the incoming partons. 
The charge and angular momentum are  
rapidly shed during the `balding' phase leaving a Schwarzschild BH which 
then decays by Hawking radiation described by the Hawking temperature, $T_H$. 
(It is assumed that passing through the balding phase does not alter the 
applicability of the cross section formula above by more than factors of order 
unity.) Other global thermodynamic quantities such as the entropy, $S$, 
specific heat/heat capacity, $C$, 
and (scaled) free energy, $F=(M_{BH}-T_H S)/M_*$, are 
also useful in describing the detailed nature 
of the BH. In the EH case these quantities depend solely on $n$ and 
$M_{BH}$ and have rather simple behaviors. 
For completeness we note that if only the EH term is present one finds the 
following `standard' relationships between these quantities which are 
simple monotonic functions of $R$ or $M_{BH}$:
\begin{eqnarray}
m&=&cx_0^{n+1}\nonumber \\
T_0=T_{H_0}/M_*&=&{(n+1)\over {4\pi x_0}}\nonumber \\
S_0&=&{{4\pi c}\over {n+2}}x_0^{n+2}\nonumber \\
F_0&=&{m\over {n+2}}\nonumber \\
C_0&=&-4\pi c x_0^{n+2}\,,
\end{eqnarray}
where $m=M_{BH}/M_*$, $x_0=M_*R_0$ and the constant $c$ is given by
\begin{equation}
c={{(n+2) \pi^{(n+3)/2}}\over {\Gamma({{n+3}\over {2}})}}\,.
\end{equation}
Note that all the quantities on the left have been defined to be dimensionless 
through appropriate scalings by $M_*$. Also note that the constant $c$ 
differs here from that given by other authors in the 
literature{\cite {DL,GT}} due to our use of $M_*$ as the common parameter.  
The index `0' for the quantities in this equation 
labels the fact that they arise in 
conventional $4+n$-dimensional General Relativity assuming an asymptotically 
flat space and in the limit that 
corrections arising from the finite size of the extra dimensions, of order 
$\sim (R/R_c)^2$, can be neglected.  
$R_0$ is thus the Schwarzschild radius for a BH of mass $M_{BH}$ assuming 
the EH action in $D=4+n$ dimensions.

One can easily imagine that if any of the higher Lovelock terms are present 
then the conventional mass-radius-thermodynamical relations above 
can be significantly altered. Of course, first one needs to 
show that the asymptotically flat, $D$-dimensional Schwarzschild solution 
still exists when the new 
Lovelock terms are present in the action; fortunately this issue was 
dealt with long ago{\cite {Wheeler}} and such solutions are now well known. 
Since that 
time there have been many discussions in the literature about the properties 
of BH in Lovelock-extended gravity{\cite {big}} but very little attention 
has been given to how such terms may influence TeV-scale BH and 
the production of such BH at colliders{\cite {Barrau}}. After some work the 
analyses given in Refs.{\cite {Wheeler,big}} allow us to extract the 
complete expressions for the various thermodynamic quantities above for 
TeV-scale BH based on our generalized action. 
Our goal is to be able to input the BH mass, \ie, $m$, as well 
as the value of $n$ and the Lovelock parameters($\alpha,\beta,\gamma$) 
and then extract values for all of the other quantities of interest. 

The most important relation is the one between $M_{BH}$ and 
the Schwarzschild radius, $R$ in this more general case; one obtains:
\begin{eqnarray}
m&=&c\Big[x^{n+1}+\alpha n(n+1)x^{n-1}+\beta n(n+1)(n-1)(n-2)x^{n-3}
\nonumber \\
 &+&\gamma n(n+1)(n-1)(n-2)(n-3)(n-4)x^{n-5}\Big]\,,
\end{eqnarray}
where $x=M_*R$ and the `0' index has now been dropped. 
Again, we remind the reader that the term proportional to 
$\gamma(\beta,\alpha)$ is only present if $n\geq 5(3,1)$. 
Of course what we really want to know is $x(m)$ so that we 
must find the roots of this polynomial equation; this must be done 
numerically in general except for some special cases. Fortunately, for 
the parameters of interest to us below, we find that  
this polynomial has only one distinct real positive root. The roots of this 
equation do lead  to some remarkable properties (that will be 
discussed in full detail below). To whet our appetite, consider 
a simple situation 
with $n=3$ but where $\alpha=\gamma=0$; in this case, the equation above 
can be inverted trivially. Then, for physical values of $x$ it is clear that 
$m$ must be bounded from below, \ie, unless the mass is in excess of a 
certain minimum, no horizon will form. This leads to a 
remarkable threshold-like 
behavior in the BH cross section that we will return to below. 

Once $x(m)$ is known, the BH temperature can be obtained as usual from 
the metric. The asymptotically flat 
$n$-dimensional BH metric in the presence of the Lovelock 
terms has the general form{\cite {Wheeler}} 
\begin{equation}
ds^2=e^{2\mu}dt^2-e^{-2\mu}dr^2-r^2 d\Omega^2\,,
\end{equation}
with $e^{2\mu}\to 1$ as $r\to \infty$. The dimensionless 
BH temperature is then given by{\cite {Wheeler,big}}
\begin{equation}
T={1\over {4\pi}} (e^{-2\mu})'|_S\,,
\end{equation}
with the prime here denoting differentiation with respect to `r' and the index 
implying evaluation at the Schwarzschild radius, so that one obtains 
\begin{equation}
T={(n+1)\over {4\pi}}{U(x)\over {V(x)}}\,,
\end{equation}
where
\begin{eqnarray}
U(x)&=&x^6+\alpha n(n-1)x^4+\beta n(n-1)(n-2)(n-3)x^2\nonumber \\
    &+&\gamma n(n-1)(n-2)(n-3)(n-4)(n-5)\nonumber \\
V(x)&=&x\Big[x^6+2\alpha n(n+1)x^4+3\beta n(n+1)(n-1)(n-2)x^2\nonumber \\
    &+&4\gamma n(n+1)(n-1)(n-2)(n-3)(n-4)\Big]\,. 
\end{eqnarray}
The BH 
entropy can then be calculated using the familiar thermodynamical relation 
\begin{equation}
S=\int_0^x dx ~T^{-1} {\partial m\over {\partial x}}\,,
\end{equation}
which yields 
\begin{eqnarray}
S&=&{{4\pi c}\over {n+2}}\Big[x^{n+2}+2\alpha (n+1)(n+2)x^n+3\beta n(n+1)(n+2)
(n-1)x^{n-2}\nonumber \\
&+&4\gamma n(n+1)(n+2)(n-1)(n-2)(n-3)x^{n-4}\Big]\,. 
\end{eqnarray}
Here we have followed the work in Ref.{\cite {big}} and required the entropy 
to vanish for a zero horizon size. Using the expressions above the (scaled) 
free energy can now be calculated via the relation $F=m-TS$ while the heat 
capacity/specific heat is given by
\begin{eqnarray}
C&=&{{\partial m} \over {\partial T}}={{\partial m} \over {\partial x}}
\Bigg[{{\partial T} \over {\partial x}}\Bigg]^{-1}\nonumber \\
 &=&-4\pi cx^{n-6} ~V \Big[V'/V-U'/U \Big]^{-1}\,,  
\end{eqnarray}
with $U,V$ given above and with the prime now denoting differentiation with 
respect to $x$. 

\begin{figure}[htbp]
\centerline{
\includegraphics[width=8.5cm,angle=90]{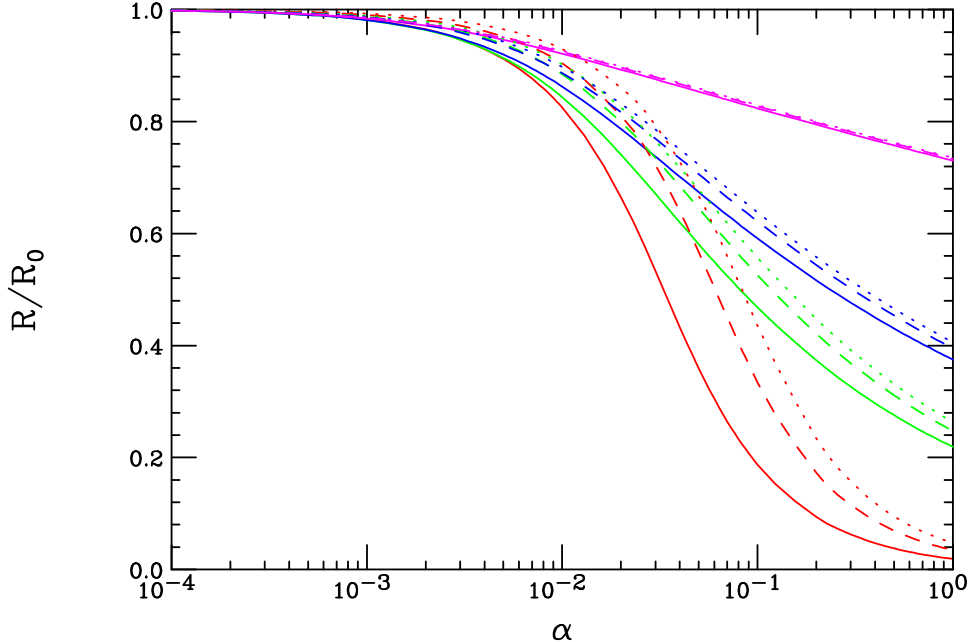}}
\vspace*{0.1cm}
\caption{Influence of $\alpha \neq 0$ on the BH mass-Schwarzschild radius 
relationship for $n=2,4,6$ and 20, corresponding to the red, green, blue and 
magenta sets of curves, respectively. The solid, dashed and dotted curves in 
each case correspond to $m=M_{BH}/M_*=2,5,8$, respectively. Here 
$\beta=\gamma=0$.}
\label{fig1}
\end{figure}

Knowing the Lovelock generalizations of all of these quantities we can now 
proceed with our analysis. We will see that significant departures from the 
expectations of the EH action are easily obtainable. While examining these 
various quantities it is important to recall that which BH properties are 
measurable within the ADD framework. As mentioned above, collider 
measurements of other ADD signatures will tell us the values of $n$ and 
$M_*${\cite {JM}}, while the BH mass itself can be determined by direct 
reconstruction{\cite {Harris,Tanaka}}; similarly 
the cross section, \ie, $R$ can 
be obtained directly. Given $M_{BH},n$ and $M_*$, various BH decay 
distributions will help to determine other quantities such as the 
temperature{\cite {Harris,Tanaka}}. Measurement of the average 
multiplicity in the BH decay can give us a handle on the BH temperature as 
$<N>\simeq m/2T${\cite {DL}} (before fragmentation effects are included). 
Of course, quantities such as the entropy and free energy are not directly 
measurable but are of theoretical interest. The specific heat of the BH 
can be inferred in some cases as it is indirectly 
probed by the BH lifetime. As we will 
see below, in contrast to the usual EH scenario, the BH may now have a 
positive specific heat making it grow cooler as it sheds mass via Hawking 
radiation. In such a case the BH may have a long lifetime. 

\begin{figure}[htbp]
\centerline{
\includegraphics[width=8.5cm,angle=90]{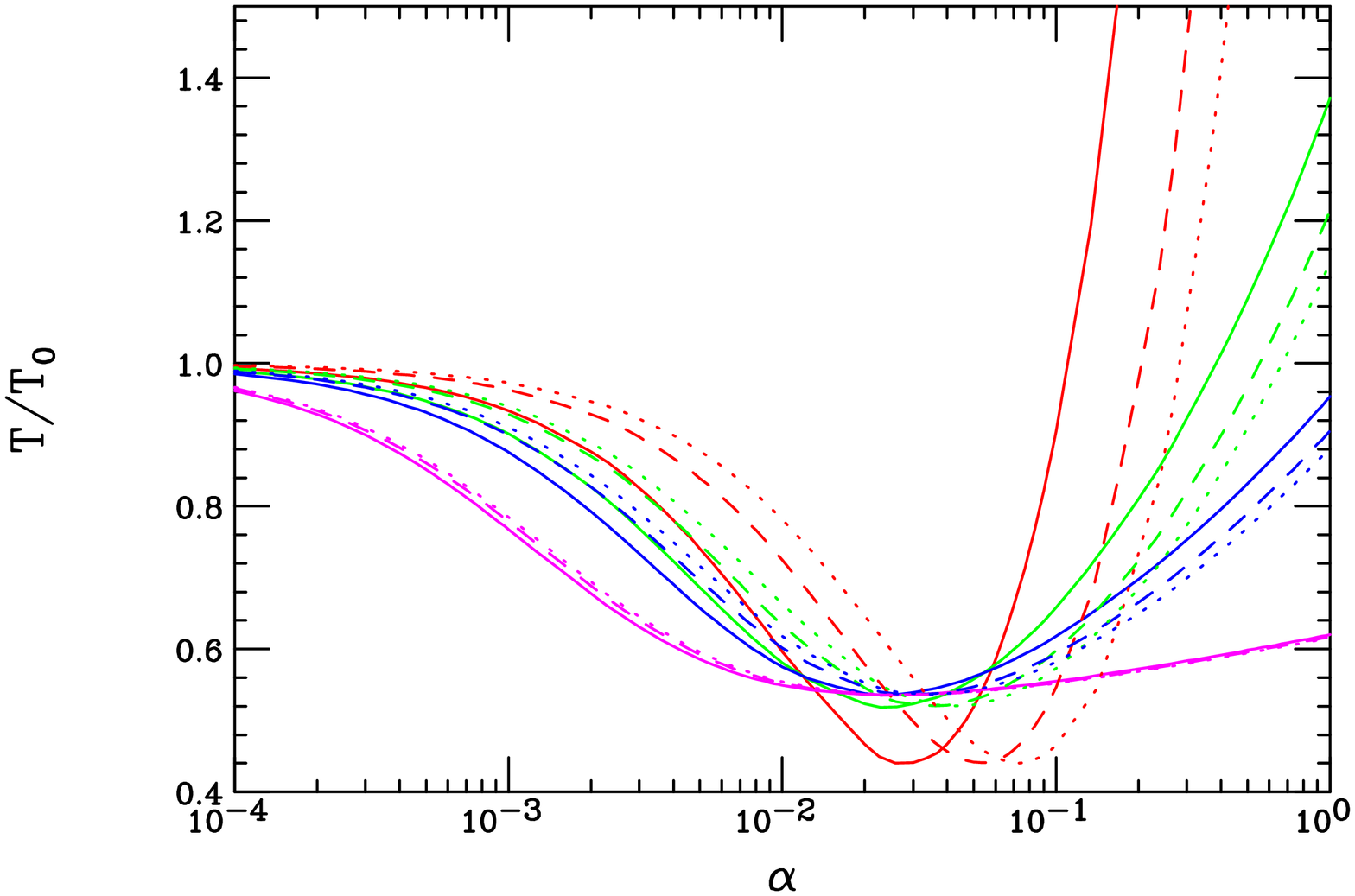}}
\vspace{0.4cm}
\centerline{
\includegraphics[width=8.5cm,angle=90]{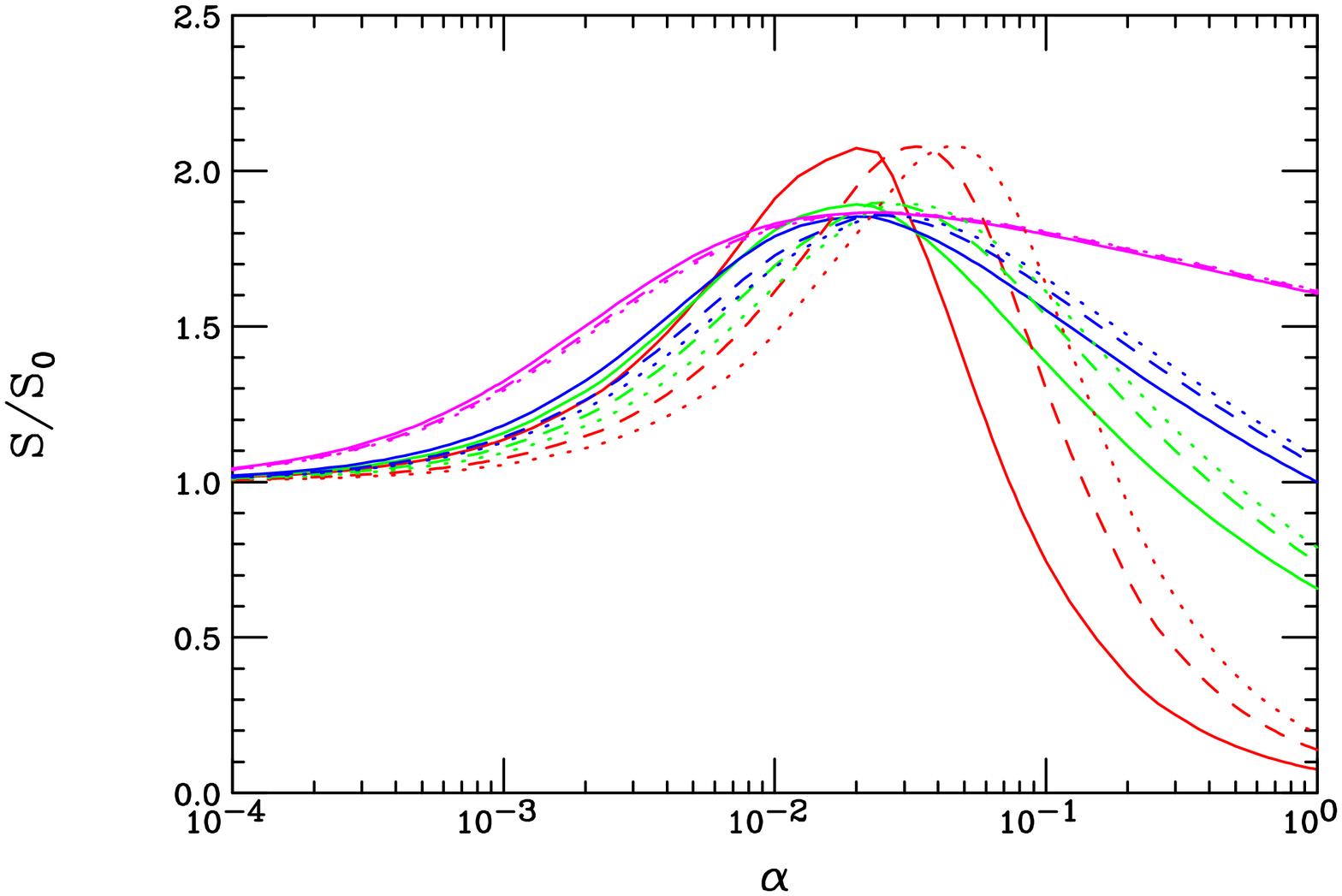}}
\vspace*{0.1cm}
\caption{Same as in the previous figure but now for the BH temperature(top) 
and entropy(bottom).}
\label{fig2}
\end{figure}

\section{Analysis}

Given the generalized expressions above we can now ask how much of a 
numerical influence the Lovelock invariants will have on the quantities of 
interest. The first issue to 
address is the values of the parameters $\alpha,\beta$ and $\gamma$. As a 
first guess one might naively expect these to be of order unity. However, 
{\it if} we think of the Lovelock terms as arising from a perturbative-like 
expansion of the full 
action, as in string theory, the coefficients must grow smaller for the higher 
order Lovelock terms. Also, when we examine the expressions above we see 
that for any kind of perturbative expansion to make sense we must have, at 
least approximately, $\alpha n^2,\beta n^4$ and $\gamma n^6<1$. Since $n$ can 
be as large as 6 within the usual ADD scenario, very 
crudely, we might expect that $\alpha \sim 10^{-2}$, 
$\beta \sim 10^{-3}-10^{-4}$ and $\gamma \sim 10^{-5}$ with a 
wide margin allowed for errors in these estimates. 
Since our arguments point to potentially different values of interest, in 
what follows we will allow for significant ranges of these 
Lovelock parameters.

\begin{figure}[htbp]
\centerline{
\includegraphics[width=8.5cm,angle=90]{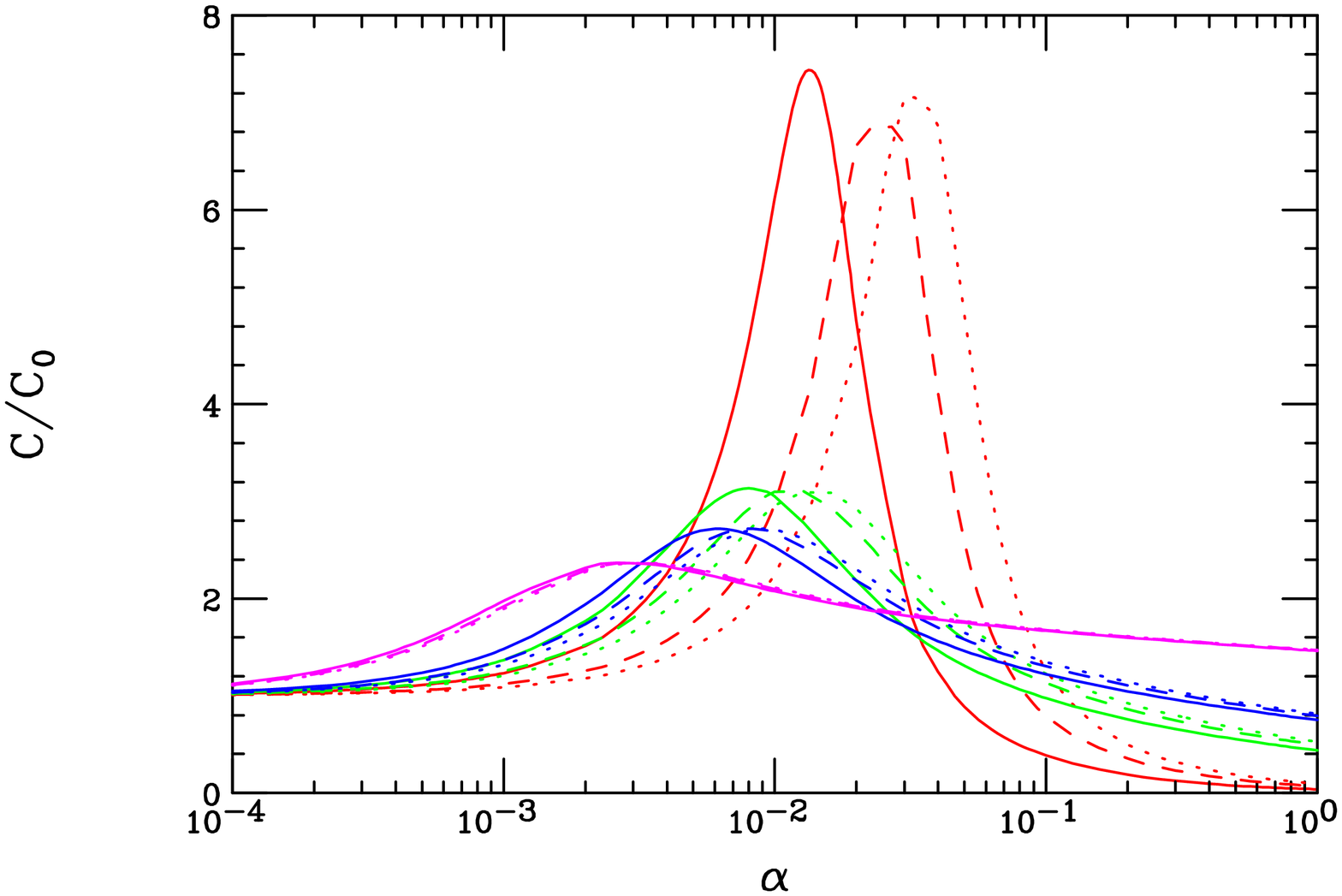}}
\vspace{0.4cm}
\centerline{
\includegraphics[width=8.5cm,angle=90]{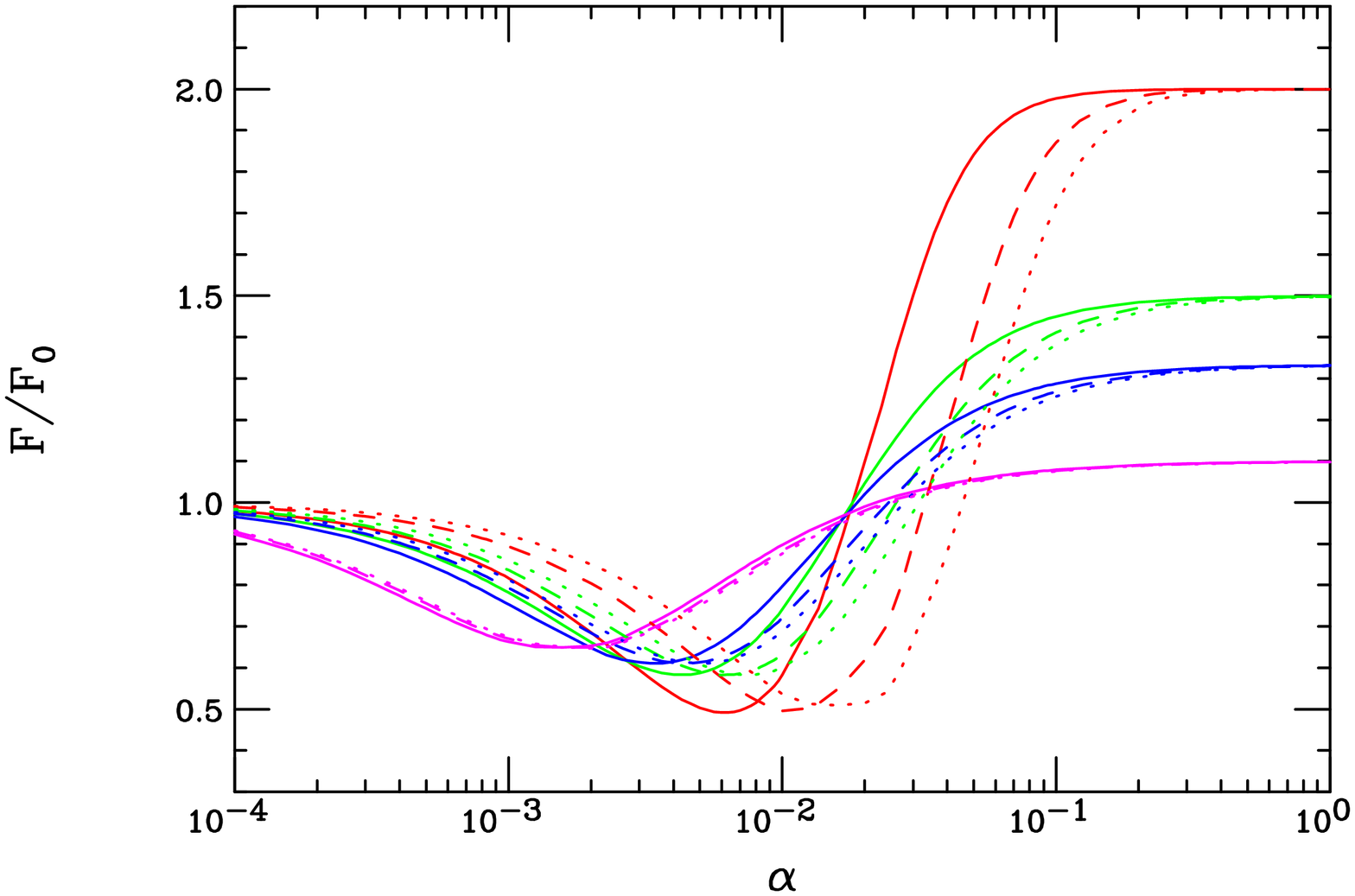}}
\vspace*{0.1cm}
\caption{Same as in the previous figure but now for the BH heat capacity(top) 
and free energy(bottom).}
\label{fig3}
\end{figure}

In order to simplify our analysis of the influence of the various Lovelock 
invariants it is perhaps best to examine them one at a time; we first set 
$\beta=\gamma=0$ and consider the case of $\alpha \neq 0$. This is the most 
familiar possibility corresponding to Gauss-Bonnet gravity discussed in the 
literature {\cite {Wheeler,big}}. In order to elucidate the effects of 
$\alpha \neq 0$ for any fixed value of $n$ and $M_{BH}$ it is useful to 
scale out the results 
by the predictions of the usual EH action as can be done using the 
expressions above. As we will see the various quantities of interest will 
have a reasonably simple dependence on $\alpha$ for fixed values of 
$m=M_{BH}/M_*$. For example, in the specific 
case of the Schwarzschild radius we will examine the ratio $R/R_0$ to which 
we now turn and which is displayed in Fig.~\ref{fig1}. Here we see that the 
effect of a non-zero $\alpha$ on $R$ is rather mild for all values of $n$ 
unless $\alpha$ is larger than $\sim 10^{-2}$. Interestingly, on the right 
hand side of the plot we see that the cases with 
larger values of $n$ appear to be significantly less influenced by a non-zero 
$\alpha$ than do those with smaller $n$. These variations in $R$ can lead to 
significant modifications to the anticipated BH cross section. 
For example, if $\alpha \simeq 1$ 
and $n=2(6)$ the BH production cross section, 
$\sim R^2$, would be reduced by more than a  
factor of 100(5) due to the presence of this new Lovelock in the action in 
comparison to EH expectations.  

\begin{figure}[htbp]
\centerline{
\includegraphics[width=8.5cm,angle=90]{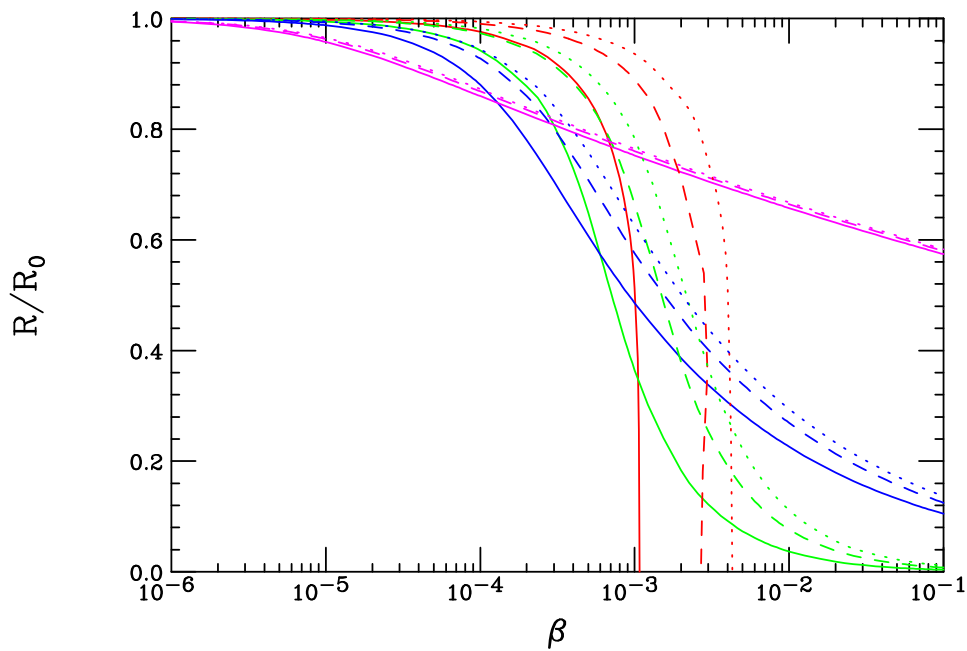}}
\vspace*{0.1cm}
\caption{Same as Fig.1 but now as a 
function of $\beta$ with $\alpha=\gamma=0$ and $n=2 \to n=3$ since 
${\cal L}_3$ vanishes when $n=2$.}
\label{fig4}
\end{figure}

Fig.~\ref{fig2} shows the corresponding results for the temperature and 
entropy ratios, $T/T_0$ and $S/S_0$, as functions of $\alpha$. The results 
shown 
in these two plots appear to be reasonably (anti-)correlated as a result of 
of the entropy definition, Eq.(15). In the top panel we see that $T/T_0$ 
initially decreases as $\alpha$ increases, then goes through a minimum near 
$\alpha \sim 0.03-0.08$, then increases with a rapidity which is highly $n$ 
dependent. We see that as $n$ increases the steepness of this temperature 
rise rapidly decreases. Again, larger $n$ appear to be less sensitive to 
appreciable values of $\alpha$. The reverse behavior is seen in the case of 
the entropy variation with both $\alpha$ and $n$. Note that in both cases the 
ratios never differ from unity by more than factors of a few unless $\alpha$ 
becomes quite large. Thus, qualitatively, these BH are not too dissimilar 
from those of EH except for extreme choices of the parameter $\alpha$.

\begin{figure}[htbp]
\centerline{
\includegraphics[width=8.5cm,angle=90]{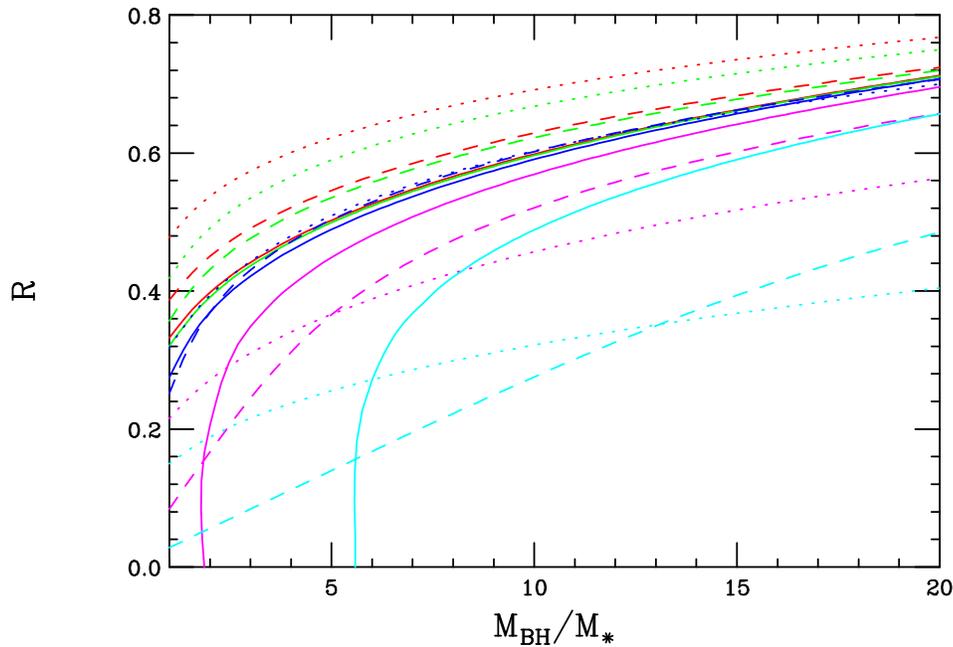}}
\vspace*{0.1cm}
\caption{BH mass dependence of the Schwarzschild radius for the case 
$\alpha=\gamma=0$. Solid(dashed,dotted) curves correspond to $n=3(4,6)$, 
respectively. The red(green,blue,magenta,cyan) curves correspond to 
$\beta=0.00003(0.0001,0.0003,0.001,0.003)$, respectively. Note here that $R$ 
is in units of $M_*^{-1}$.}
\label{fig5}
\end{figure}

In Fig.~\ref{fig3} we see the behavior of both the heat capacity and free 
energy ratios with variations in $\alpha$. The values of these two quantities 
are reasonably simple and qualitatively similar to both $T$ and $S$. 
$F/F_0$ has an easily understandable 
behavior with increasing $\alpha$ since $F=m-TS$ and given the behavior  
above of the ratios $T/T_0$ and $S/S_0$. It is interesting to note that at 
large $\alpha$ the ratio $F/F_0$ asymptotes to a set of values which depend on 
$n$ and not $m$. As in the EH case, $\partial F/\partial m$ can be shown 
to be always positive and the 
heat capacity for BH is seen to be well behaved and is always negative as 
in the EH case. Thus 
BH with only non-zero $\alpha$ will Hawking radiate, evaporating completely 
or until a Planck-scale remnant stage is reached. For very large values of 
$\alpha \sim 1$, the mass loss through evaporation may be somewhat slowed 
in comparison to the usual EH case since the BH heat capacity, 
$\partial m/\partial T$, is so small. From Figs. 1-3, however,   
it is generally apparent that when $n\geq 2$ and 
only $\alpha$ is non-zero the typical BH will 
behave in a manner which is qualitatively similar to (though quantitatively 
different from) those that arise solely from the EH action. They are all 
unstable and will have relatively short lifetimes typical of more familiar 
hadronic processes.

\begin{figure}[htbp]
\centerline{
\includegraphics[width=8.5cm,angle=90]{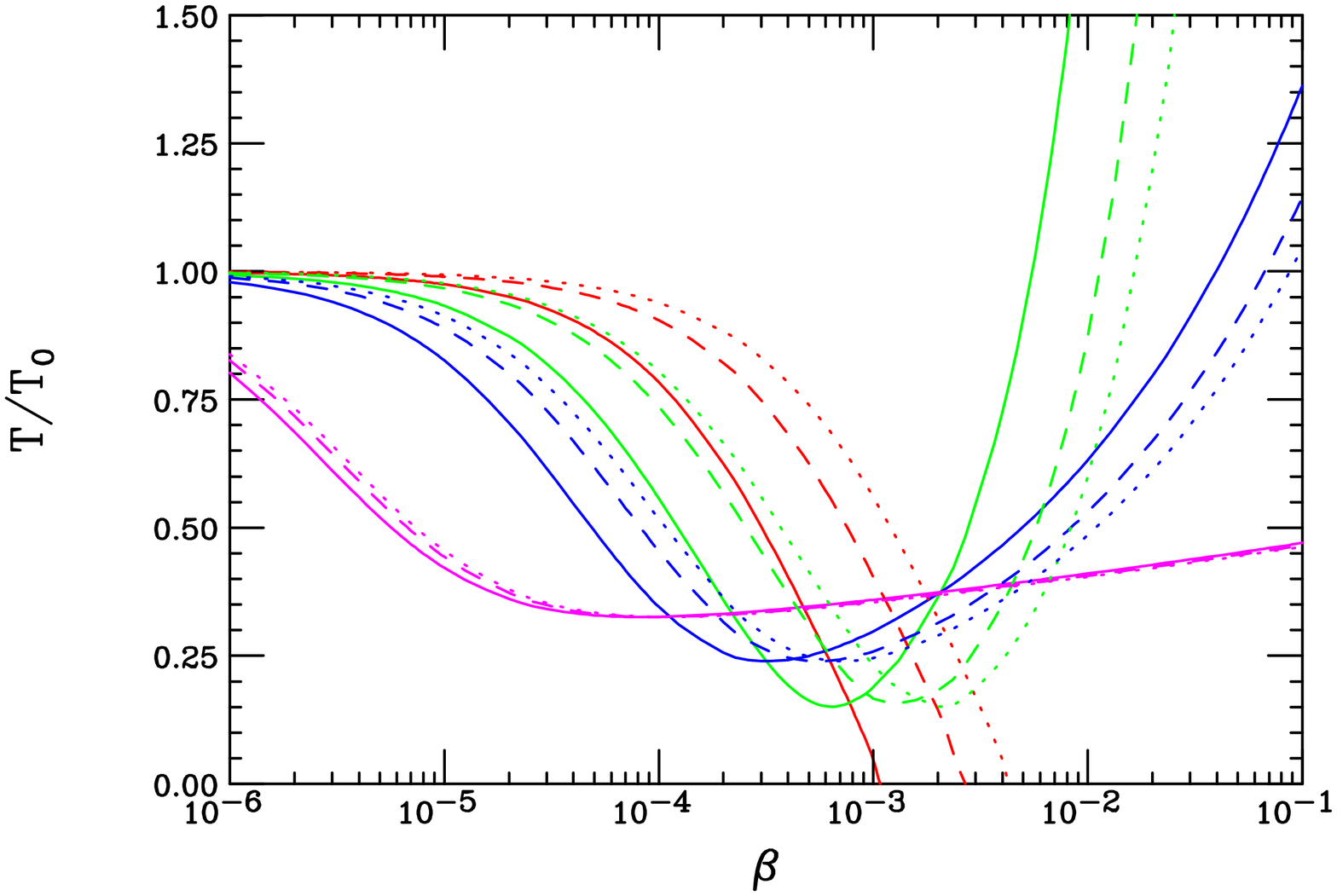}}
\vspace{0.4cm}
\centerline{
\includegraphics[width=8.5cm,angle=90]{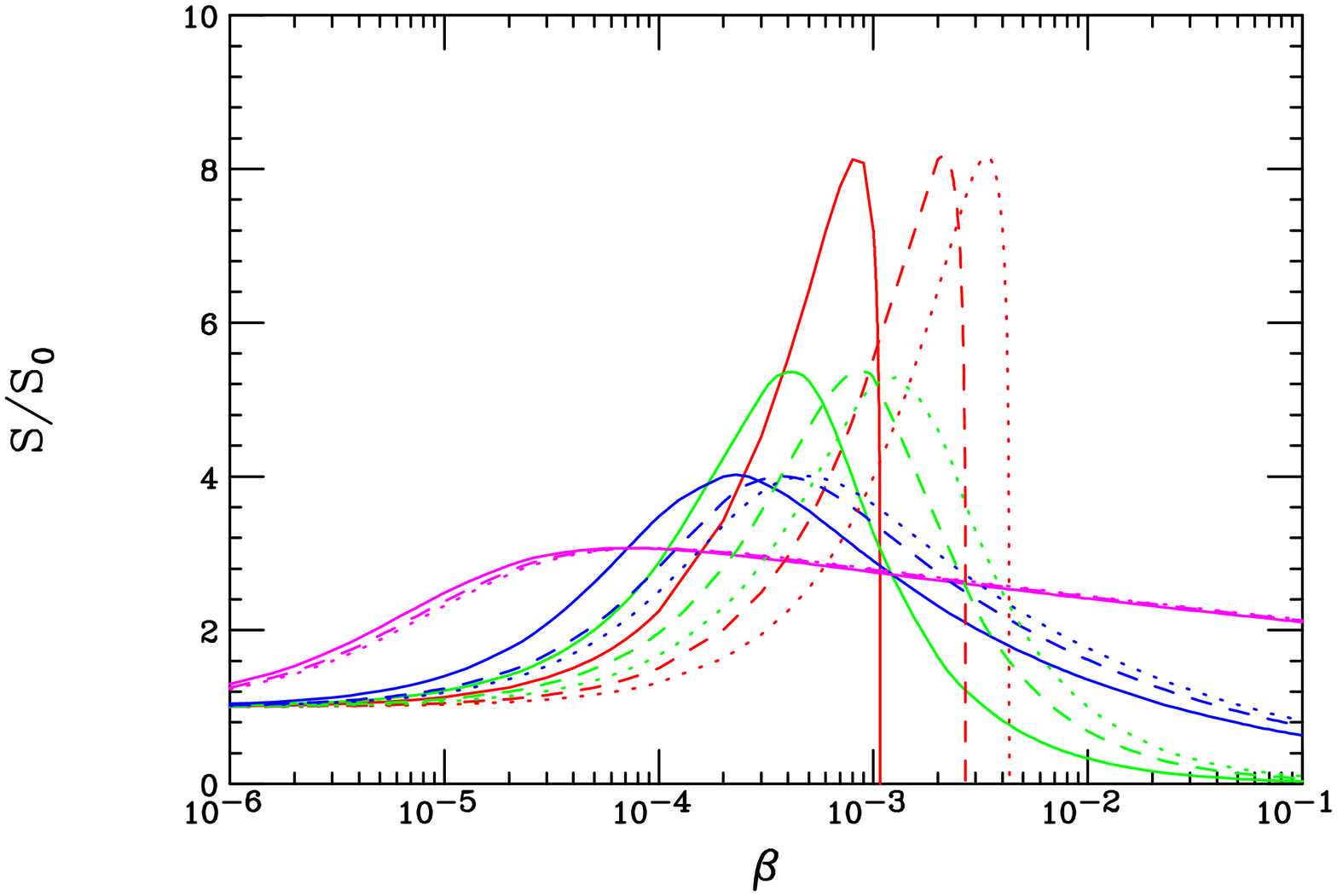}}
\vspace*{0.1cm}
\caption{Same as Fig.2 but now as a function of 
$\beta$ with $\alpha=\gamma=0$ and with $n=2 \to n=3$.}
\label{fig6}
\end{figure}

The behavior of BH properties with non-zero $\beta$ can be significantly 
different than what happens in the case of non-zero $\alpha$ as 
we will now see. (For simplicity we will assume $\alpha=\gamma=0$ while 
studying non-zero $\beta$; we note, however, that the behavior of non-zero 
$\gamma$ is qualitatively similar to non-zero $\beta$.) 
We can see this as soon as we examine the $\beta$ 
dependence of the ratio $R/R_0$ as is shown in Ref.~\ref{fig4}. For small 
values of $\beta$ we see that the ratio $R/R_0$ decreases away from unity 
with smaller values of $n$ most seriously affected. This is quite similar 
to what we saw happen in the the case of non-zero $\alpha$. 
Clearly either large $\alpha$ 
or large $\beta$ can lead to a significant suppression of the BH production 
cross section; the same will be true of 
non-zero $\gamma$. Here, however, we see something 
new for the case $n=3$: the ratio $R/R_0 \to 0$ for finite $\beta$. Turning 
this around, this means that for $n=3$ a non-zero $\beta$ implies that an 
horizon will not form unless $m$ exceeds a certain critical value, \ie, 
a BH will not form unless a certain mass threshold is reached. (Such a 
possible behavior has already been observed within the general context of 
Lovelock extended actions{\cite {Wheeler,big}}). This threshold effect can be 
seen analytically from Eq.(10) by setting $n=3$ with $\alpha=\gamma=0$ which 
yields the simple relation
\begin{equation}
m=c(x^4+24\beta)\,,
\end{equation}
where here $c=5\pi^3/2$ and which is trivial to invert analytically to
\begin{equation}
x=\Big[{m\over {c}}-24\beta\Big]^{1/4}\,.
\end{equation}
Here we see that unless $m>60\pi^3 \beta$ there is no horizon. Note that for 
$\beta=5\cdot 10^{-4}$, a typical value, this means that unless 
$m>0.93$ no BH are formed. 
It is important to observe that a similar result is {\it not} obtainable 
when $n>3$; the case $n=3$ is therefore 
special when $\beta \neq 0$. Note that for fixed $\beta$ simultaneously 
turning on a non-zero $\alpha$ will lead to this same qualitative 
behavior with the same minimum value of $m$. Perhaps even more interesting 
is the fact that something almost identical 
happens in the case of non-zero $\gamma$ when $n=5$.

\begin{figure}[htbp]
\centerline{
\includegraphics[width=8.5cm,angle=90]{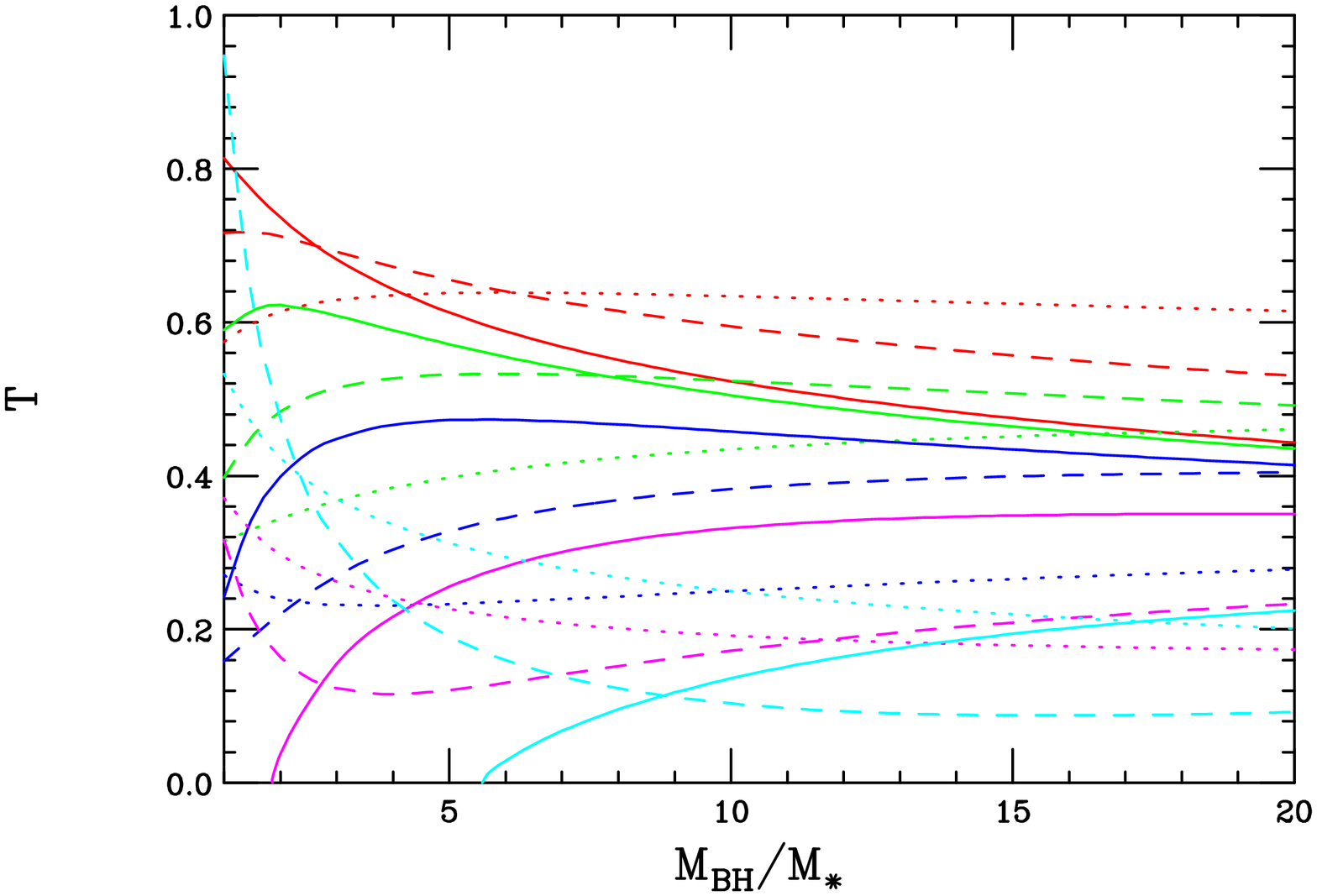}}
\vspace{0.4cm}
\centerline{
\includegraphics[width=8.5cm,angle=90]{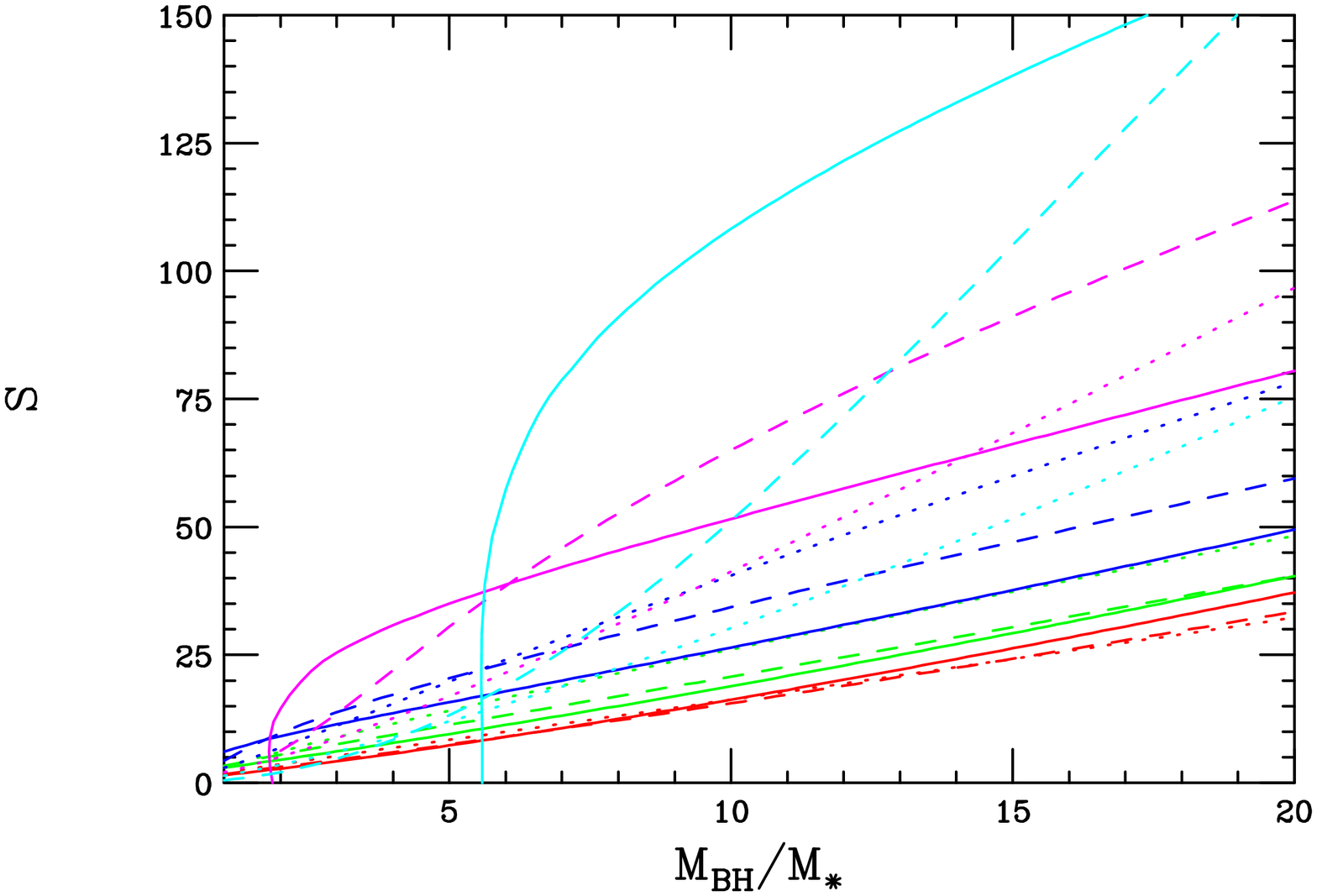}}
\vspace*{0.1cm}
\caption{Same as Fig. 5 but now for the BH temperature and entropy with 
$n=2 \to n=3$ as above. Note that 
here the temperature is in units of $M_*$.}
\label{fig7}
\end{figure}

This horizon threshold effect for non-zero values of $\beta$ can perhaps be 
best seen in Fig.~\ref{fig5} where we show $R$ as a function of 
$m=M_{BH}/M_*$. Here it is quite evident that there is a finite horizon 
threshold for the case $n=3$ while none exists for other $n$ values, \ie, 
$R\to 0$ when $m \to 0$. For 
$n>3$ this is easily understood from Eq.10 as there we see that $m=0$ 
when $x=0$. It is interesting to note that this threshold effect is not too 
dissimilar from what may happen for a 4-d BH when a renormalization 
group running 
of Newton's constant is employed{\cite {Bonanno}} in order to approximate 
leading quantum corrections; such a threshold scenario can also be seen to 
occur in theories with a minimum length{\cite {cav}} and in loop quantum 
gravity{\cite {loop}}, which may also be 
thought of as a quantum effect. It is interesting to note that this 
threshold phenomena occurs in all these models where one tries to incorporate 
quantum corrections in some way; it would be interesting to learn whether or 
not this is a general feature of all such models.

\begin{figure}[htbp]
\centerline{
\includegraphics[width=8.5cm,angle=90]{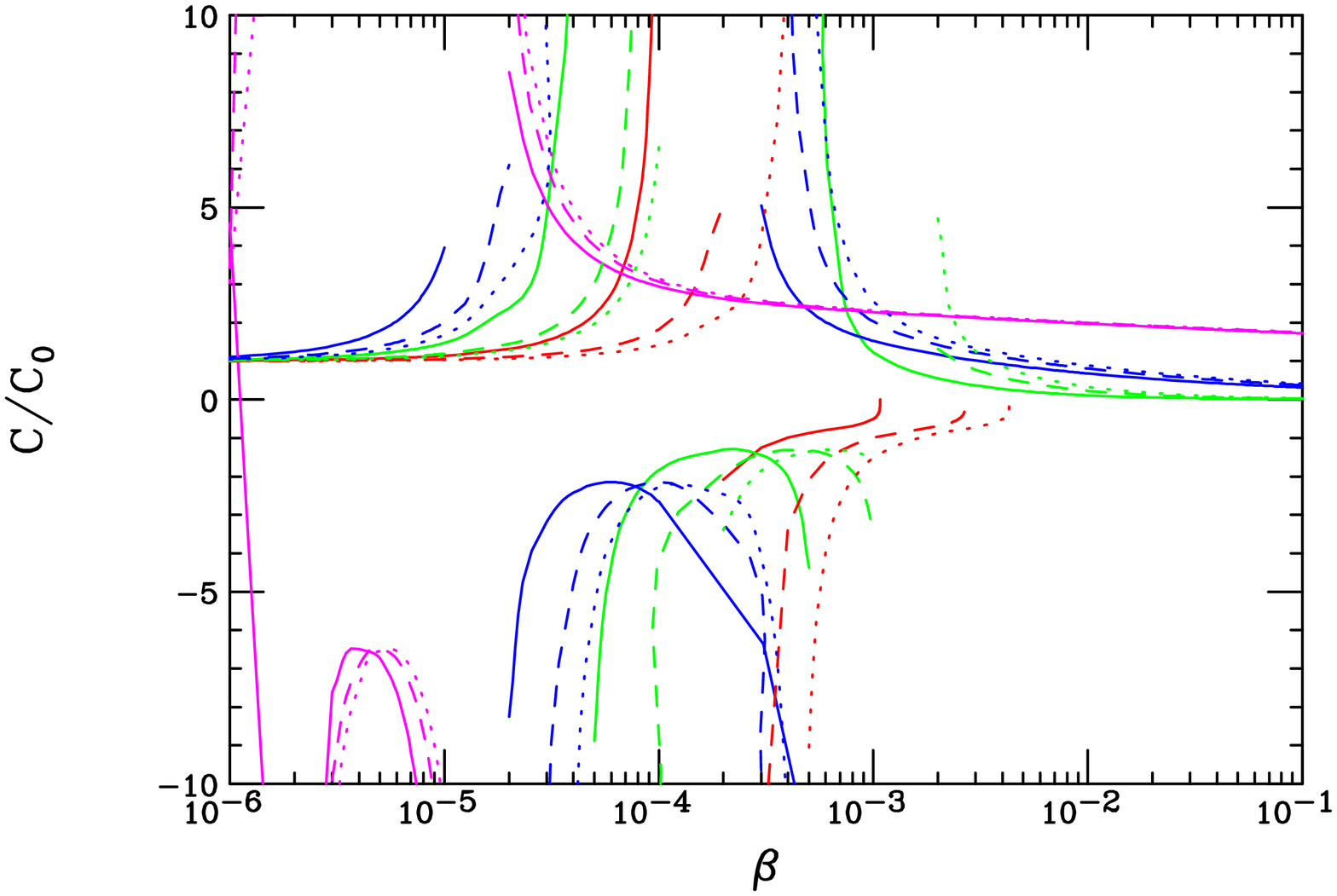}}
\vspace{0.4cm}
\centerline{
\includegraphics[width=8.5cm,angle=90]{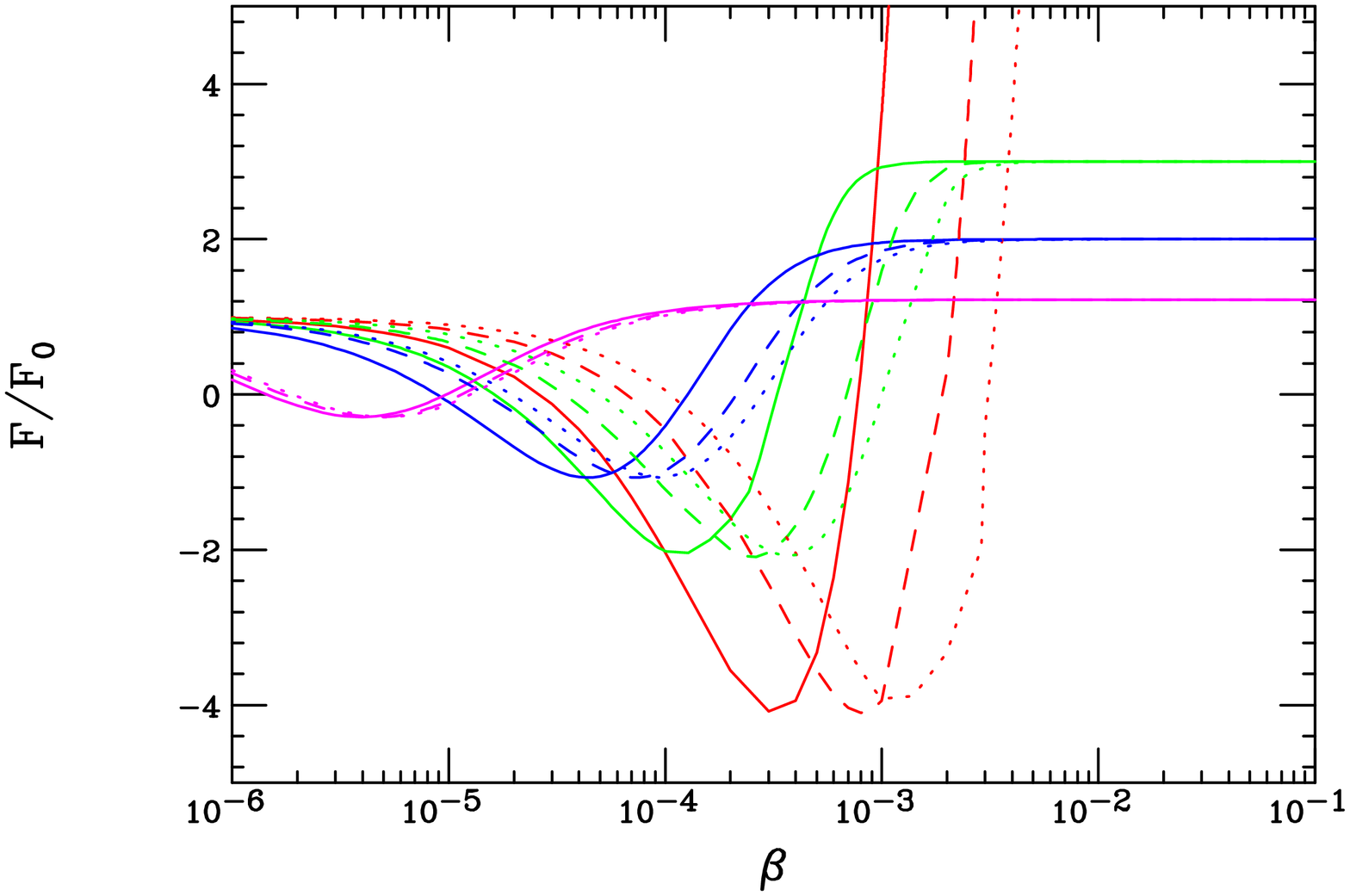}}
\vspace*{0.1cm}
\caption{Same as Fig.3 but now as a function of $\beta$ with 
$\alpha=\gamma=0$.}
\label{fig8}
\end{figure}

We might expect that this unusual threshold-type behavior will have a 
significant influence on thermodynamic quantities. Fig.~\ref{fig6} 
shows the dependence of $T/T_0$ and $S/S_0$ as functions of $\beta$ for 
different values of $n$ and $m$. For $n>3$ this behavior is not so 
qualitatively different than that seen in the case of a non-zero $\alpha$ 
which is consistent with what was observed in Figs.~\ref{fig4} and 
~\ref{fig5}. However, in Fig.~\ref{fig6} we see that for $n=3$ both the 
temperature and entropy can 
vanish. These zeros occur precisely at the values at 
which $R\to 0$ for the $n=3$ case that we observed above: $T$, $S$ and $R$ 
all vanish at the same parameter space points.  These zeros in both the 
temperature and entropy for finite $m$ are also observed for the case of 
$n=3$ in Fig.~\ref{fig7} but we also see something new for other values 
of $n$. While $\partial S/\partial m$ was always positive, it is possible 
for $\partial T/\partial m$ to change sign in the mass range of interest. 
In the EH case, as well as for the examples 
of non-zero $\alpha$ examined above, 
$\partial T/\partial m$ was always seen to be 
negative. In many cases for $\beta \neq 0$, 
$T(m)$ is first observed to rise, reach a maximum, then decrease as 
$m$ grows larger. 

\begin{figure}[htbp]
\centerline{
\includegraphics[width=8.5cm,angle=90]{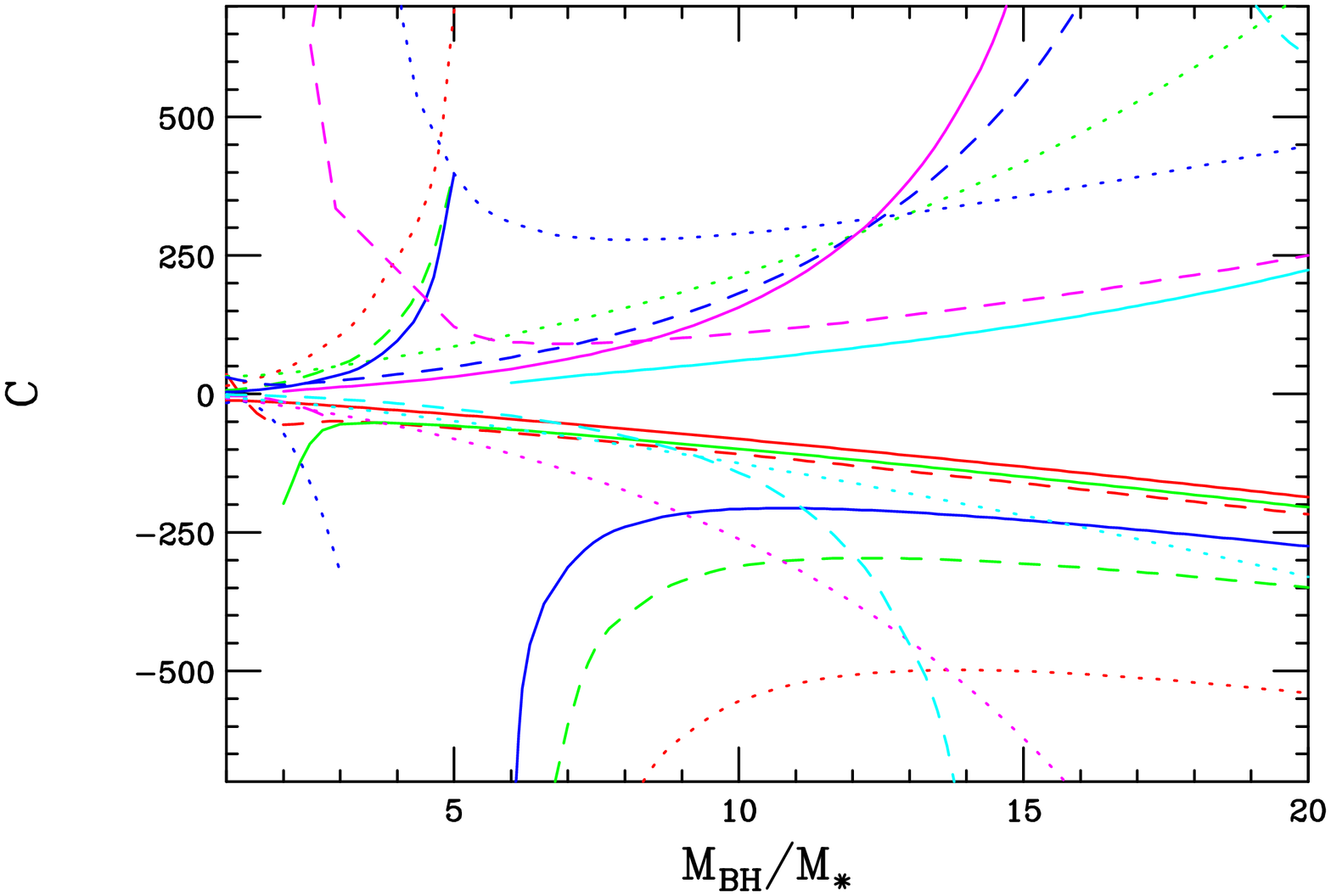}}
\vspace{0.4cm}
\centerline{
\includegraphics[width=8.5cm,angle=90]{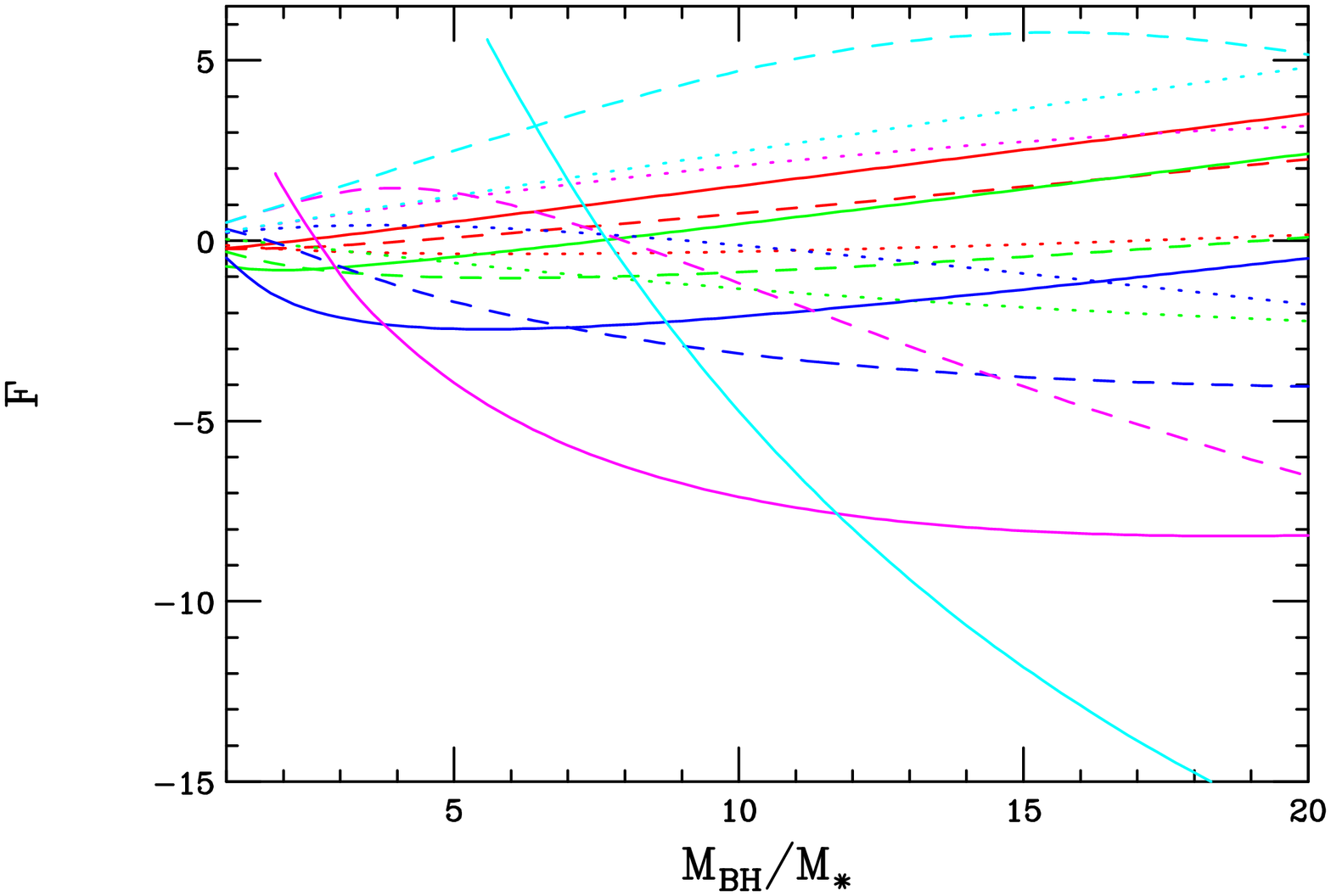}}
\vspace*{0.1cm}
\caption{Same as the Fig.7 but now for the BH heat capacity and free energy. 
Recall that in our notation $F$ is units of $M_*$ while the heat capacity is 
dimensionless. Curves terminating (and not changing sign) 
in the middle of the plot correspond to locations where $R\to 0$.}
\label{fig9}
\end{figure}

To understand this more fully we turn to Figs.~\ref{fig8} and 
~\ref{fig9} which show the BH heat capacity and free energy both of which can 
now have rather  complex behavior. The least unusual of the two is 
the ratio $F/F_0$, which apart from the case of $n=3$, 
is qualitatively similar to the that seen above for the case 
of a non-zero $\alpha$. However, we still see that $F$ can become negative 
for some values of $\beta$. On the otherhand, the specific 
heat/heat capacity is 
observed to have as many as two sign reversals associated with singularities 
in the parameter range of interest. In most cases $C/C_0$ is positive for 
both large and small $\beta$ with the reversals taking place at intermediate 
values. After some algebraic manipulations, it can 
be shown that for fixed values of $m$, $C(\beta)/C_0$ generally 
has two simple poles for real positive $\beta$ values. Such behavior is 
not unheard of for BH arising from Lovelock extended 
actions{\cite {Wheeler,big}} and may even occur in 4-d{\cite {Lousto}} 
for Kerr-Neumann BH. 
For the $n=3$ case, however, since the range of $R$ and $m$ is 
restricted there can be only one singularity.

\begin{figure}[htbp]
\centerline{
\includegraphics[width=8.5cm,angle=90]{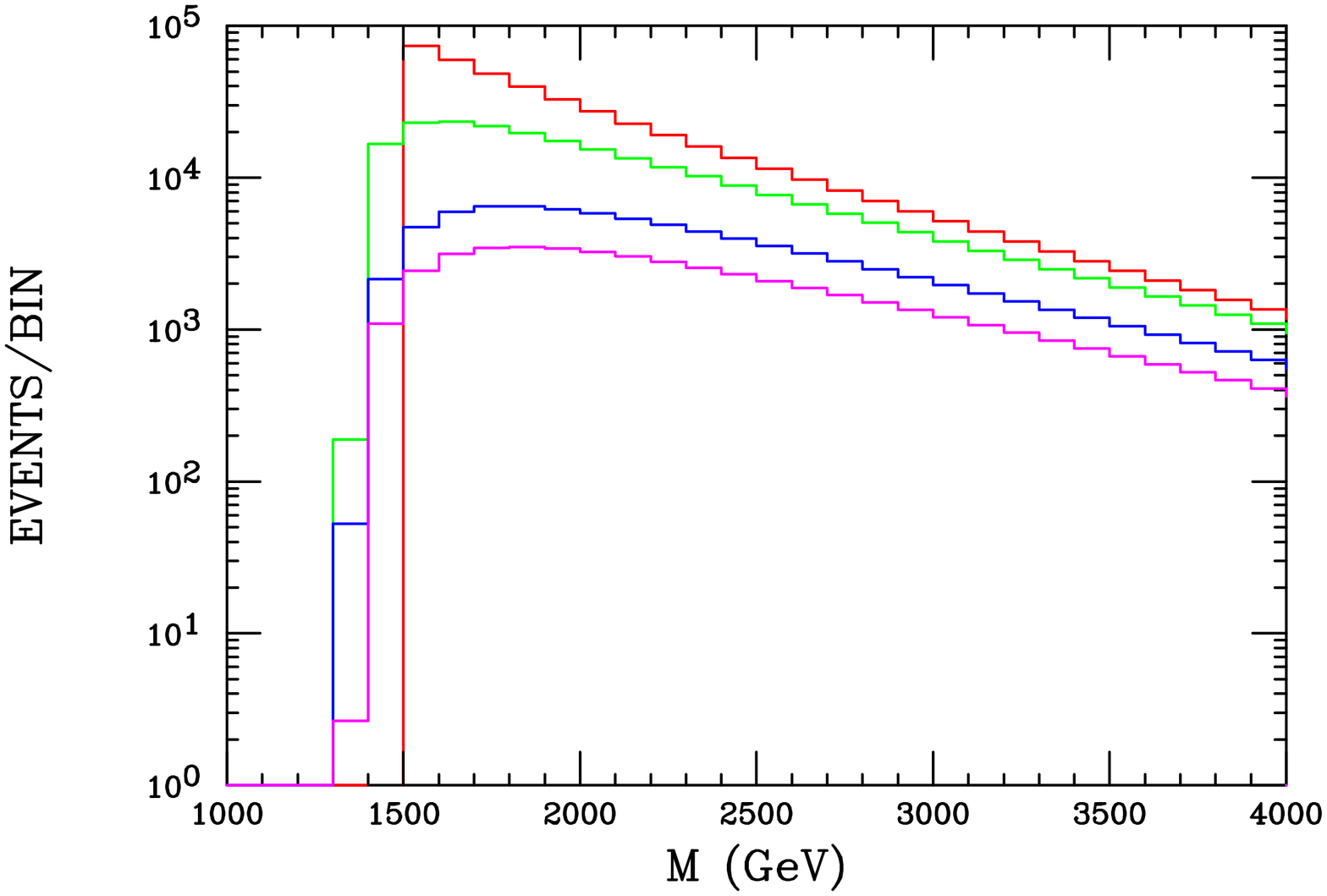}}
\vspace{0.4cm}
\centerline{
\includegraphics[width=8.5cm,angle=90]{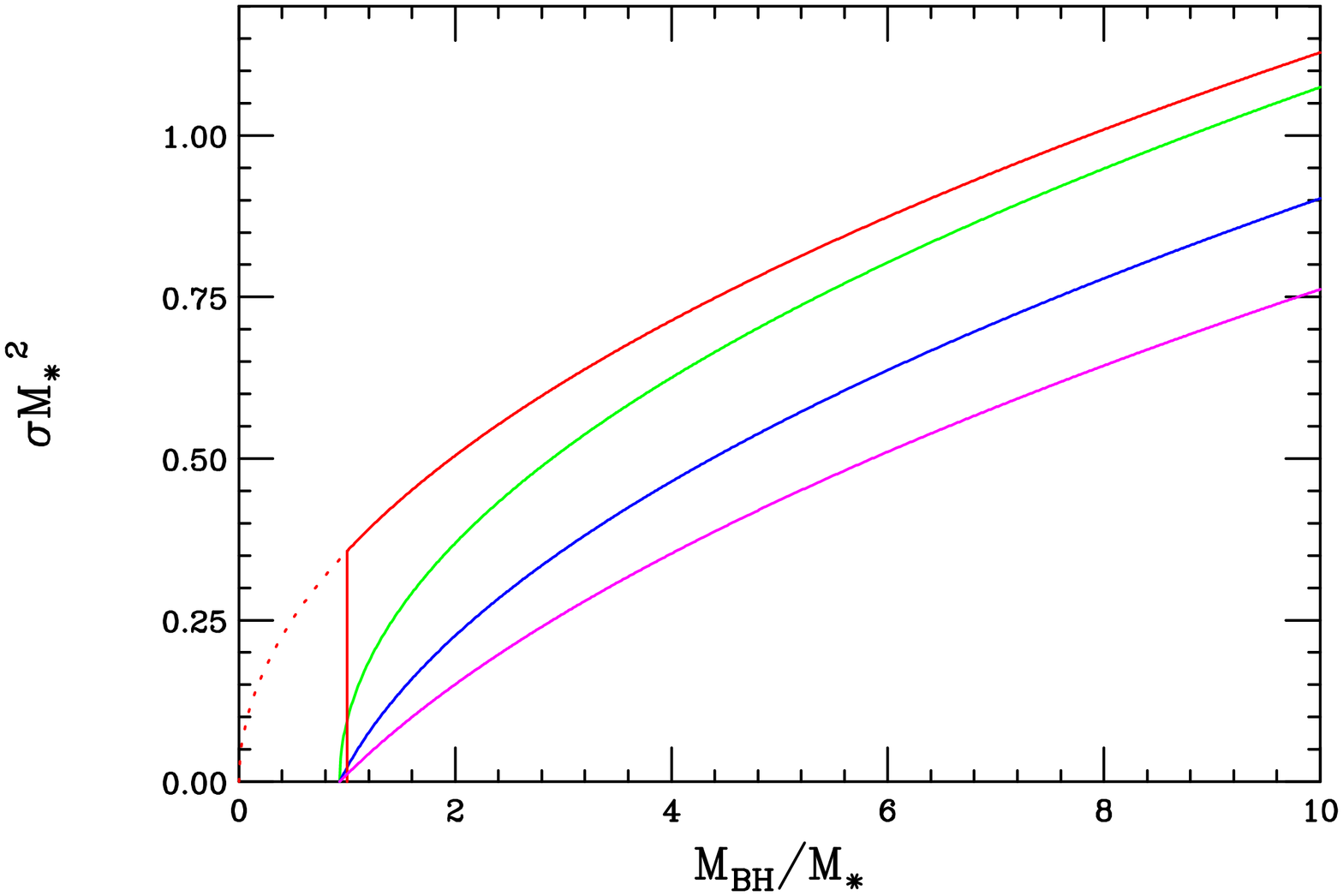}}
\vspace*{0.1cm}
\caption{Threshold behavior of the BH cross section at the LHC(top) and 
ILC(bottom) for $n=3$. The red curve assumes the absence of Lovelock terms, 
while the green(blue,magenta) one corresponds to $\beta=0.0005$ with 
$\alpha=0(0.01,0.02)$. In the LHC case a luminosity(bin width) of 100 
fb$^{-1}$ (100 GeV) has been assumed. In the ILC case a scaled cross section 
is presented as discussed in the text; the usual step-function turn on in 
the absence of Lovelock terms has been assumed. For the ILC $\sqrt s=M_{BH}$.}
\label{fig10}
\end{figure}

If we instead examine $C(m)$, as shown in Fig.~\ref{fig9}, we see that there 
is at most 
one singularity corresponding to a simple pole in the physical region. 
For small $m$ we see that $C>0$ and then flips sign as $m$ increases 
above the singularity.  This signals the existence of 
a kind of phase transition occurring 
at this singularity. $F(m)$ is also unusually 
behaved: in general it is positive for both small and large $m$ but 
goes through 
a negative minimum in between. Not unsurprisingly the location of the minimum 
in the free energy is the same as that of the singularity in the 
specific heat/heat capacity. 

A parameter 
regime with positive heat capacity leads to interesting properties for BH. 
With only the EH action and thus a negative $C$, a BH loses mass  
via Hawking radiation; as it does 
so it get hotter thus radiating more quickly. 
This run-away process stops only when evaporation is complete 
or a non-classical 
Planck scale remnant is reached. Here, with $\beta \neq 0$, it is 
possible that the BH will end up in a mass regime where $C>0$ in which case 
it will get {\it cooler} as it gets lighter. Furthermore, there is a 
maximum temperature, given by the location of the pole in the heat capacity, 
for the BH. One might expect that such BH will live much longer than naively 
expected based on the EH action or even the combined EH+${\cal L}_2$ action. 
This is something 
completely new that happens when higher order Lovelock invariants are 
present. In some cases this may imply that the BH will become classically 
stable which is what happens in the case of $n=3$. To see this analytically, 
it is interesting to calculate the BH lifetime (in a vacuum!) directly 
assuming the semi-classical approach is viable. Recall that the amount 
of energy emitted by a black body per unit time is proportional to both its 
area and the fourth power of its temperature;  
using the results presented in Ref.{\cite {Barrau}} 
for $n=3$ with $\alpha=\gamma=0$ as an example we obtain  

\begin{eqnarray}
{dm\over {dt}}&\sim& R^2T^4\nonumber \\
          &\sim& {{(m-60\pi^3\beta)^{7/2}}\over {(m+120\pi^3 \beta)^4}}\,.   
\end{eqnarray}
Integrating from some arbitrary initial mass, $m_0$, down to the 
threshold value, $60\pi^3 \beta$ (where $T=0$),  
we obtain an infinite lifetime for all  
non-zero values of $\beta$. By `infinite lifetime' we mean that it will take 
an infinitely long time for the mass of the BH to reach the threshold value. 
These mass thresholds may be identifiable with the anticipated Planck-scale 
remnant. 
We note that if $\beta=0$ a finite lifetime would be obtained.   
The addition of a non-zero value of $\alpha$ will 
not change this infinite lifetime result in a qualitative way. Thus any BH 
with a mass above the threshold value will decay by Hawking radiation quite 
rapidly losing most of its `excess' mass until its mass approaches the 
threshold. Interestingly, we see from the figures that as the 
BH loses mass and approaches the threshold mass 
value from above the free energy approaches a minimum which is 
another indicator of classical stability. 
Of course, taking $\beta \to 0$ one recovers the finite 
lifetime that is usually expected on the basis of the EH action. 
This apparent BH stability is of course not unique to these 
particular parameter values; other BH do not lead to finite mass final states  
but may take an infinite time to evaporate completely. 
Furthermore, for $n=5$, we find that a similar situation 
is seen to occur when $\gamma$ is non-zero for arbitrary positive values of 
$\alpha,\beta$. Thus the so-called BH remnant in these cases can be uniquely 
identified with a BH with a mass very close to the threshold value.

\begin{figure}[htbp]
\centerline{
\includegraphics[width=8.5cm,angle=90]{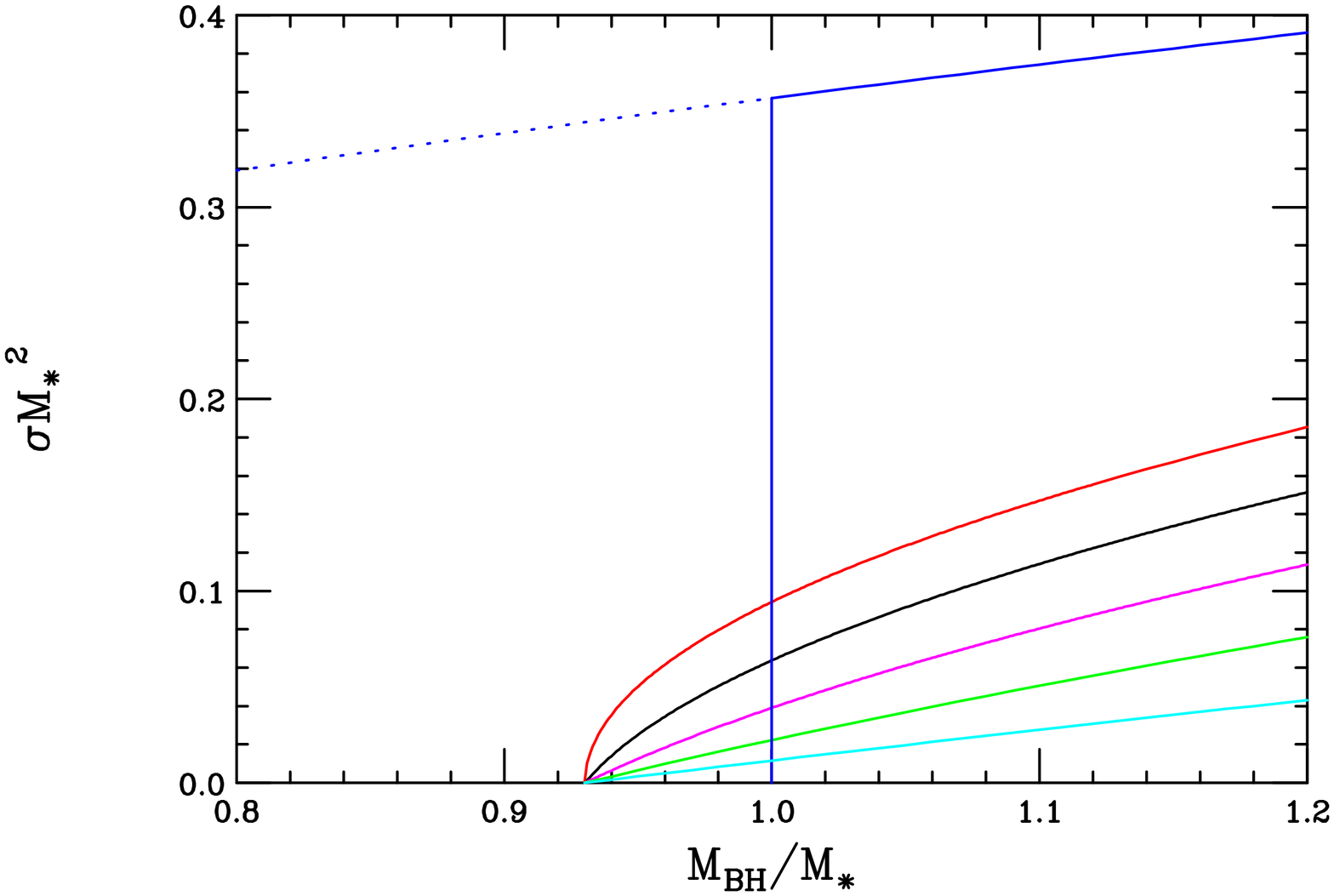}}
\vspace{0.4cm}
\centerline{
\includegraphics[width=8.5cm,angle=90]{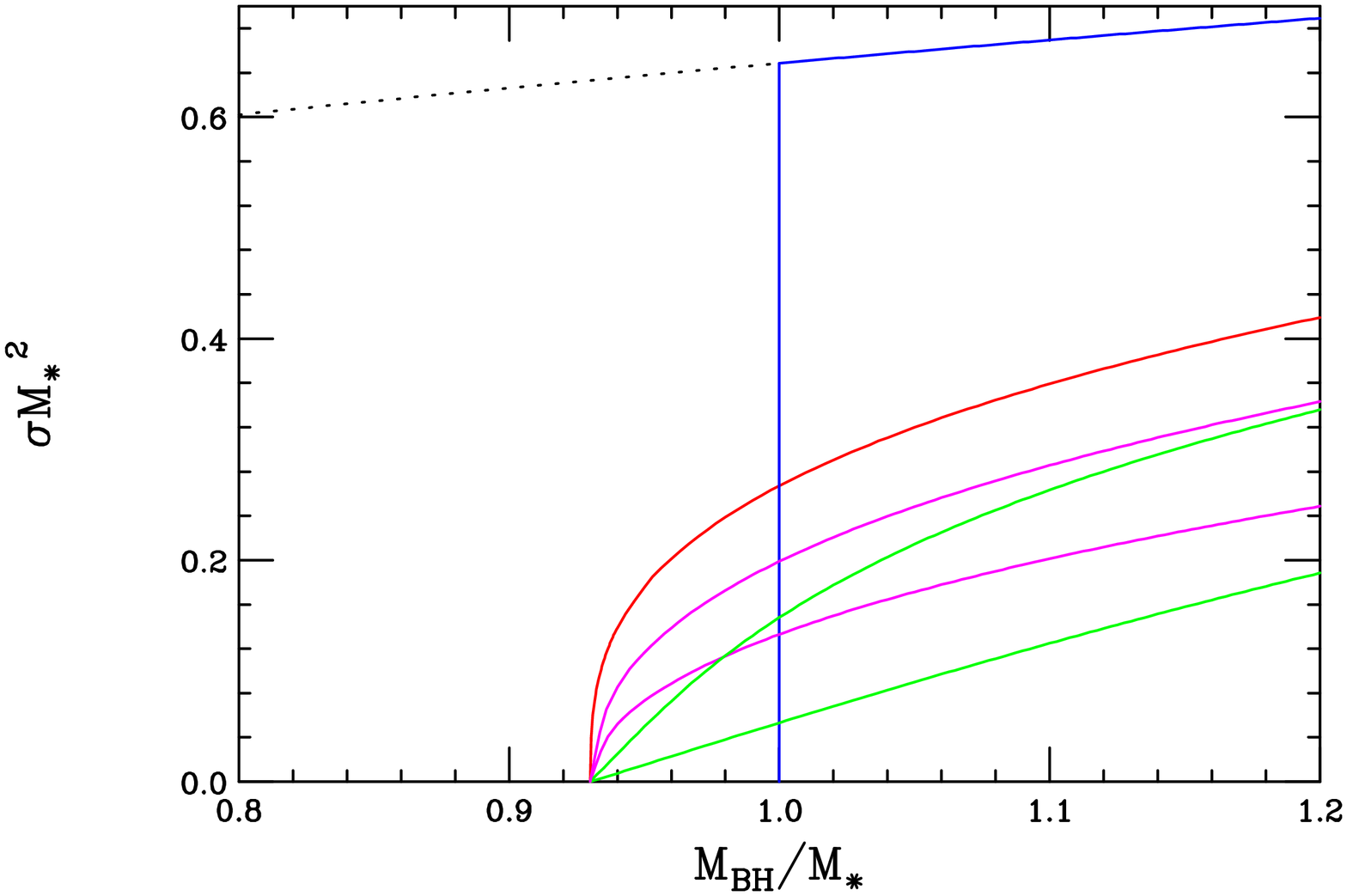}}
\vspace*{0.1cm}
\caption{Close up of the BH production threshold at the ILC for $n=3$(top) 
and $n=5$(bottom); in these plots we should remember that $\sqrt s=M_{BH}$. 
The blue line in the naive $\theta$-function that is 
usually assumed in the absence of Lovelock terms. For the case $n=3$ 
with $\beta=0.0005$ the 
curves correspond to $\alpha=0,0.002,0.005,0.01$ and 0.02 from top to bottom, 
respectively. For the case $n=5$ all curves have 
$\gamma \simeq 1.14\cdot 10^{-5}$ and, from top to bottom, have 
$\alpha,\beta=0,0(0.003,0; 0,0.00003; 0.01,0; 0,0.0001)$, respectively.}
\label{fig11}
\end{figure}

At this point it is important that the reader be reminded that even 
though we are trying to model quantum/stringy corrections via the Lovelock 
invariants, these are still  semi-classical predictions which are subject 
to true quantum corrections that may 
modify them significantly. Though classically stable, these BH may have finite 
lifetimes due to these quantum corrections. It is possible, however, that 
these lifetimes will remain quite long.

\section{ A Toy Model of BH Collider Cross Section Thresholds}

As discussed above, the usual picture of collider BH production is that there 
is a $\theta$ function-type threshold at $M_*$ which is certainly 
unphysical. Clearly quantum and/or stringy corrections will turn this step 
function behavior into something smooth which might be modeled via additional 
Lovelock terms in the action. If TeV scale extra dimensions exist and BH are 
produced at future colliders, 
the threshold region, where these effects 
are most relevant, will be important to understand in detail since it will 
provide a handle on the true underlying theory of quantum 
gravity. In the scenario 
we propose below, this corresponds to extracting the values of the Lovelock 
parameters by examining the nature of the cross section threshold. This is 
akin to using the nature of the threshold for new particle production to 
determine the particle's spin and other quantum numbers. 

The analysis in the previous section suggests a toy model that 
yields a more properly behaved threshold for BH production, at least in 
the cases $n=3(5)$, when $\beta(\gamma)$ is 
non-zero. There we saw that, \eg, a 
non-zero $\beta$ for $n=3$ implies the existence of a minimum mass below which 
a BH simply will not form. We also saw that this mechanism was absent for even 
values of $n$. For reasonable values of the Lovelock parameters 
what do these BH thresholds look like?
Sticking with the case 
$n=3$ for now, let us take $\beta=0.0005$, in the center 
of the range discussed above with $\alpha$ set to zero for the moment for 
simplicity. Note that since $n=3$, $\gamma$ is automatically zero. The top 
panel in Fig.~\ref{fig10} shows what this would imply for the LHC BH (very 
large) production 
rate assuming that $M_*=1.5$ TeV for purposes of demonstration. 
First we notice the red histogram which is the usual $\theta$-function turn-on 
cross section prediction.  The green histogram shows the same cross section 
but now with $\beta=0.0005$ and no $\theta$-function. 
There is clearly a threshold near 1.5 TeV but the turn-on is no longer 
infinitely sharp; for large BH masses we see that this prediction 
asymptotes to that of the naive 
$\theta$-function histogram. This result we may have expected since for large 
values of $M_{BH}$ we should recover the predictions associated with the 
EH action. Turning on non-zero but typical values of $\alpha \sim 0.01$ 
does not alter the threshold location but does modify the cross section 
shape near the threshold. Apparently the larger the value of $\alpha$ the 
softer the slope of the cross section near threshold. In fact a short 
analysis shows 
that the slope of the cross section goes as $\sim 1/\alpha$ 
precisely at the threshold. Of course, radiative corrections will smear out 
the fine details of these 
rather sharp threshold shapes. For non-zero $\alpha$ the cross 
section is seen to still asymptote to the naive value in the large $M_{BH}$ 
limit as it should. 

Although a detailed simulation is far beyond the scope of the present work it 
is clear from the top panel in Fig.~\ref{fig10} that, while the LHC will 
observe the BH production threshold, an extraction of the Lovelock parameter 
values with any precision will be difficult. The bottom panel of this figure 
shows what may be seen at the ILC; here a set of scaled cross section results 
are presented so that the reader can move the threshold as they prefer. 
For example, if  
$M_*=1$ TeV, $\sigma M_*^2=1$ corresponds to a cross section of $\simeq 389.4$ 
pb; there will be no shortage of statistics. As in the case of the LHC the 
red curve is the naive step function turn-on while the other curves show the 
effect of the non-zero Lovelock parameters. To get a better idea of the ILC  
threshold region we zoom in using the top panel of Fig.~\ref{fig11}. In this 
idealized picture we see that the step function curve is very far away from 
any of those produced by the Lovelock terms. The Lovelock cross section 
predictions  are themselves rather widely separated from each other and have 
quite distinct shapes as they approach the threshold. Again, we recall 
that if $M_*=1$ TeV, a single vertical 
tickmark on this plot corresponds to a cross 
section change of $\sim 8$ pb, \ie, $\sim 10^6$ events at the ILC in one year. 
Of course in the real world this threshold 
region will be made more complex by both 
electroweak and gravitational radiative corrections that may smear out the 
threshold to some extent. Just how significant such effects will be is not 
yet known even within the very limited scope of the present toy model. 

For the case of $n=5$ a non-zero $\gamma$ also induces a BH mass threshold 
and the bottom panel shows the corresponding cross sections assuming 
$\gamma \simeq 1.1\cdot 10^{-5}$, a value in the anticipated 
region discussed above. 
Note that the slope of the cross section in the threshold region 
is generally increased as $n$ increases. Turning on different values of 
$\alpha$ and $\beta$ again leads to detailed modifications of the 
threshold cross section shape. When $\alpha=\beta=0$, the slope of the cross 
section at threshold is infinite; a short analysis shows that in this case the 
slope behaves as $\sim 1/\beta$ near threshold and is independent of $\alpha$. 
The various curves are seen to 
be sufficiently different as to be likely distinguishable at the ILC even 
after radiative effects are included. 
Thus we see that the cases $\beta \neq 0$ 
and $\gamma \neq 0$ are qualitatively similar but are clearly separable once 
measurements with any precision become possible.

At least for the case of $n$ odd we see that the Lovelock invariants can lead 
to a reasonably smooth turn on of BH production in the mass range near $M_*$. 
The precise nature of this turn in is highly correlated with the various 
parameter values and the number of extra dimensions.

\section{Conclusions}

In this paper we have examined how the production of TeV-scale black holes 
at colliders is influenced by the existence of higher curvature Lovelock 
invariants in the action of models with large extra dimensions. This 
possibility is anticipated on rather general grounds when we try to embed 
such models in a larger theoretical structure, \eg, strings, if we also wish 
to maintain unitarity and freedom from ghosts. 
In the case of the ADD model, we argued that 
the addition of the Lovelock invariants to the Einstein-Hilbert action 
would leave the more conventional missing-energy and spin-2 contact 
interaction signatures unaltered but would modify the anticipated properties 
of TeV scale BH since the ADD geometry is essentially flat. These 
modifications in BH properties provide the {\it only} signature for the 
existence of higher curvature terms in this context. Since the total number of 
dimensions $D=4+n$ in string-motivated ADD is $\leq 10$ and the properties of 
the Lovelock invariants are highly restrictive, it was easily shown that at 
most three such terms, ${\cal L}_{2,3,4}$, can  
be added to the usual Einstein-Hilbert action. The strengths of these new 
interactions are described by a set of dimensionless parameters and, 
given the expectation that such terms result from some form of perturbative 
expansion, we derived some rough estimates for their anticipated magnitudes. 

The next step in our analysis was to obtain the general expressions 
for the Schwarzschild 
radius, Hawking temperature, entropy, heat capacity/specific heat and free 
energy relevant for the $D$-dimensional TeV-scale BH in Lovelock gravity 
models with our specific generalized action. Using 
these results we then performed a detailed numerical investigation 
to probe the sensitivity of the BH production cross section as well as 
these thermodynamic quantities to variations 
in the values of the Lovelock parameters. Since the number of potentially 
contributing invariants is $n$-dependent, we examined the possibility 
of non-zero values for coefficients ($\alpha$ and $\beta$) of both 
${\cal L}_{2,3}$ one at a time; while the impact of ${\cal L}_2$ and 
${\cal L}_3$ were found to be quite distinct, ${\cal L}_4$ leads to results 
which are quite similar to ${\cal L}_3$ and so was not examined in detail. 

For non-zero values of $\alpha$, up to factors or order unity, the BH 
thermodynamical properties are overall 
qualitatively similar to those found in the pure 
EH case except when $\alpha$ get very large. The BH Schwarzschild radius is, 
however, significantly smaller which suppresses the collider BH production 
cross section. When $\beta$ is non-zero, both quantitative as well as 
qualitative variations in the BH properties 
can be anticipated especially for the case of 
$n=3$; these variations from EH expectations can be somewhat exotic. We showed 
that they include sign changes in the heat capacity/specific heat, indicative 
of phase transitions, negative free energies and BH mass thresholds, \ie, mass 
values below which BH cannot form. In particular we observed that BH can now 
become thermodynamically stable, \ie, it would take 
an infinite time for them to 
Hawking radiate down to their threshold masses. This is correlated with BH now 
possibly having positive heat capacities so that they cool as they lose 
mass thus slowing the mass loss process itself. 

We then 
demonstrated that for $n=3,5$ the minimum BH masses induced by the Lovelock 
invariants could be used as a toy model for the BH production cross section 
threshold. Instead of a step function turn-on, the cross section near 
threshold was shown to be highly sensitive to the values of $\alpha,\beta$ and 
$\gamma$ yet would still asymptote to that expected in the pure EH case. These 
differences would surely show up at the LHC and can be probed in detail at the 
ILC due to the large cross section and the enormous statistics available. It 
may be possible to use these BH mass thresholds to extract the values of the 
Lovelock parameters themselves.

TeV scale BH can provide a unique tool to probe the underlying theory of 
gravity. Hopefully they will be seen at future colliders.

\noindent{\Large\bf Acknowledgments}

The author would like to thank J.Hewett and B. Lillie for discussions 
related to this work.

%
\def\MPL #1 #2 #3 {Mod. Phys. Lett. {\bf#1},\ #2 (#3)}
\def\NPB #1 #2 #3 {Nucl. Phys. {\bf#1},\ #2 (#3)}
\def\PLB #1 #2 #3 {Phys. Lett. {\bf#1},\ #2 (#3)}
\def\PR #1 #2 #3 {Phys. Rep. {\bf#1},\ #2 (#3)}
\def\PRD #1 #2 #3 {Phys. Rev. {\bf#1},\ #2 (#3)}
\def\PRL #1 #2 #3 {Phys. Rev. Lett. {\bf#1},\ #2 (#3)}
\def\RMP #1 #2 #3 {Rev. Mod. Phys. {\bf#1},\ #2 (#3)}
\def\NIM #1 #2 #3 {Nuc. Inst. Meth. {\bf#1},\ #2 (#3)}
\def\ZPC #1 #2 #3 {Z. Phys. {\bf#1},\ #2 (#3)}
\def\EJPC #1 #2 #3 {E. Phys. J. {\bf#1},\ #2 (#3)}
\def\IJMP #1 #2 #3 {Int. J. Mod. Phys. {\bf#1},\ #2 (#3)}
\def\JHEP #1 #2 #3 {J. High En. Phys. {\bf#1},\ #2 (#3)}

\end{document}